\documentclass{aa}  
\usepackage{graphicx}
\usepackage{txfonts}
\def\as{\arcsec\hspace{-1.2mm}.\hspace{0.3mm}} 
\def\am{\arcmin\hspace{-1.2mm}.\hspace{0.3mm}}

\begin{document}
   \title{The Dancing Sky: 6 years of night sky observations at Cerro
   Paranal \thanks{Based on observations made with ESO Telescopes at
   Paranal Observatory.}}

   \subtitle{}

   \author{F. Patat\inst{1}}

   \offprints{F. Patat}

   \institute{European Southern Observatory (ESO), K. Schwarzschildstr. 2,
              D-85748, Garching b. M\"unchen, Germany\\
              \email{fpatat@eso.org}
             }

   \date{Received ...; accepted ...}

% \abstract{}{}{}{}{} 
% 5 {} token are mandatory
 
  \abstract
   {}
   {The present work provides the results of the first six years of operation 
   of the systematic night-sky monitoring at ESO-Paranal (Chile).}
   {The $UBVRI$ night-sky brightness was estimated on about 10,000 
    VLT-FORS1 archival images, obtained on more than 650 separate nights, 
    distributed over 6 years and covering the descent from maximum to minimum 
    of sunspot cycle n. 23. Additionally, a set of about 1,000 low resolution, 
    optical night-sky spectra have been extracted and analyzed.}
   {The unprecedented database discussed in this paper has led to the 
    detection of a clear seasonal variation of the broad band night sky 
    brightness in the $VRI$ passbands, similar to the well 
    known semi-annual oscillation of the Na~I~D doublet. The 
    spectroscopic data demonstrate that this seasonality is common to all 
    spectral features, with the remarkable exception of the OH 
    rotational-vibrational bands. A clear dependency on 
    the solar activity is detected in all passbands and it is particularly 
    pronounced in the $U$ band, where the sky brightness decreased by 
    $\sim$0.6 mag arcsec$^{-2}$ from maximum to minimum of solar cycle n.~23.
    No correlation is found between solar activity and the intensity of the 
    Na~I~D doublet and the OH bands. A strong correlation between the 
    intensity of N~I 5200\AA\/ and [OI]6300,6364\AA\/ is reported here for the 
    first time. The paper addresses also the determination of the correlation
    timescales with solar activity and the possible connection with the flux
    of charged particles emitted by the Sun.}
   {}

   \keywords{atmospheric effects -- site testing -- techniques: photometric -- 
   techniques: spectroscopic}

   \maketitle
%
%________________________________________________________________

\section{\label{sec:intro}Introduction}

Soon after the beginning of VLT science operations in Paranal, ESO
started an automatic $UBVRI$ sky brightness survey, with the aim of
both characterizing the site and studying the long term trend, in
order to detect any possible effects of human activity.  The results
obtained during the first 18 months of operations (April 2000 -
September 2001) have been presented and discussed in Patat
(\cite{paperI}, hereafter Paper~I). This programme, which makes use of
all scientific images obtained with FORS1, is building one of the most
extensive, accurate and homogeneous optical sky brightness data sets
ever studied. As shown in Paper~I, these data allow a very detailed
analysis, including the study of correlations with other parameters,
and the investigation of short, medium and long term variations.  The
interested reader can find exhaustive reviews on this subject in Roach
\& Gordon (\cite{roach}) and in Leinert et al. (\cite{leinert}), while
references to other published sky brightness surveys are given in
Paper I.

Since the publication of the first results, obtained on 174 different
nights close to maximum of solar cycle n.~23, the data base has been
steadily growing and progressively extending towards the solar
minimum.  In this paper I present a global analysis run on the whole
data set, which includes broadband observations taken on 668 separate
nights between 20 Apr. 2001 and 20 Jan. 2007.  Additionally, I present
and discuss here a set of more than 1000 low-resolution long-slit
night sky spectra taken with FORS1 between May 1999 and
Feb. 2005. They were used to measure the fluxes of single lines or
integrated OH bands.

The paper is organized as follows. In Sec.~\ref{sec:obs} I describe
the observations and the basic data reduction steps for photometry and
long slit spectroscopy. In Sec.~\ref{sec:dark} I review the general
results obtained during dark time, while the correlation with solar
activity and the seasonal variations are discussed in
Sects.~\ref{sec:sun} and \ref{sec:season}, respectively. The
spectroscopic analysis is presented in Sects.~\ref{sec:spectra},
\ref{sec:corr} and \ref{sec:short}. Finally, in Sec.~\ref{sec:disc} I
discuss the main results and summarize the conclusions in
Sec.~\ref{sec:concl}.

\section{\label{sec:obs}Observations and Data Reduction}

The data used in this work were obtained with the FOcal Reducer/low
dispersion Spectrograph (hereafter FORS1), mounted at the Cassegrain
focus of ESO--Antu/Melipal/Kueyen 8.2m telescopes (Szeifert 2002). The
instrument is equipped with a 2048$\times$2048 pixel (px) TK2048EB4-1
backside thinned CCD and has two remotely exchangeable collimators,
which give a projected scale of 0\as2 and 0\as1 per pixel (24$\mu$m
$\times$ 24$\mu$m). According to the collimator used, the sky area
covered by the detector is 6\am8$\times$6\am8 and 3\am4$\times$3\am4,
respectively.

\subsection{\label{sec:phot}Photometry}

The photometric data set includes 10,432 images obtained in the
$UBVRI$ passbands with both collimators. The reduction procedure is
described in Paper I, to which I refer the reader for a more detailed
description, while here I only recap the basic steps. All frames are
automatically processed by the FORS pipeline, which applies bias and
flat-field correction, the latter performed using twilight sky
flats. Once the instrument signatures are removed from the images, the
sky background is estimated using the robust algorithm described in
Patat (\cite{patatII}). The photometric calibration into the
Johnson-Cousins system is then achieved using zeropoints and colour
terms derived from the observation of standard star fields
(Landolt \cite{landolt}), regularly obtained as part of the FORS calibration
plan.  Finally, the observed values are corrected to zenith using the
standard procedure (see for example Garstang \cite{garstang}) and
logged together with a number of relevant parameters.

\subsection{\label{sec:spec}Spectroscopy}

The spectroscopic data set includes a sub-sample of all long-slit
science data present in the ESO archive whose proprietary period had
expired by the time this paper has been written.  For the sake of
simplicity, I have selected only the data obtained with the standard
resolution collimator and the single-port high-gain read-out mode,
since this combination is the most used for long slit spectroscopy
with FORS1. In order to accumulate a wide data sample, I have
retrieved from the VLT archive all public spectra taken with the 300V
grism coupled with the order-sorting filter GG435 and a long slit
1\as0 wide, which is the most frequently used (676 frames).  The
wavelength range was extended down to about 3600\AA\/ in the blue by
retrieving also all spectra taken with the same setup but with no
order-sorting filter, for a total of 163 frames.  To increase the
sample, I have retrieved also all spectra obtained with the grisms
600B (143 frames) and 600R (207 frames) coupled with the order-sorting
filters OG590 and GG435, respectively. Also in these two cases the
slit was 1\as0 wide. The main characteristics of each setup are
summarized in Tab.~\ref{tab:spec}. Exposure times range from a few
minutes to one hour.

\tabcolsep 1.5mm
\begin{table}
\caption{\label{tab:spec} Main properties of the FORS1 spectral data set.
Only spectra taken with sun elevation below $-$18$^\circ$ were
included.}
\begin{tabular}{cccccc}
\hline \hline
Grism & Filter & Wav. Range & Resolution   & Dispersion       & N\\
      &        & (\AA)      & (\AA\/ FWHM) &  (\AA\/ px$^{-1}$) & \\
\hline
300V  & GG435  & 4300--8900  & 12           & 2.6 & 676\\
300V  &  --    & 3615--8900(*)  & 12           & 2.6 & 163\\
600B  & OG590  & 3650--6050  & 5.3          & 1.2 & 143 \\
600R  & GG435  & 5390--7530  & 4.5          & 1.0 & 207\\ 
\hline
\multicolumn{6}{l}{(*) Second order overlapping above 6000 \AA.}
\end{tabular}
\end{table}

All images were processed within the {\tt longslit} package of
IRAF\footnote{IRAF is distributed by the National Optical Astronomy
Observatories, which are operated by the Association of Universities
for Research in Astronomy, under contract with the National Science
Foundation.}. Due to the large amount of data and the purpose of this
work, the bias subtraction was performed using only a pre-scan
correction, while flat-fielding effects were neglected.

Wavelength calibration was achieved using a 2D solution derived from a
set of reference arc exposures. Given the procedure adopted for the
spectrum extraction, this step is mandatory, since in FORS1 the line
curvature can reach a peak-to-peak value of about 10 px. If not
accounted for, this instrumental feature would produce an apparently
significant but artificial line broadening when collapsing the 2D
spectra perpendicularly to the dispersion direction (see next
section).

\begin{figure}
\centering
\includegraphics[width=9cm]{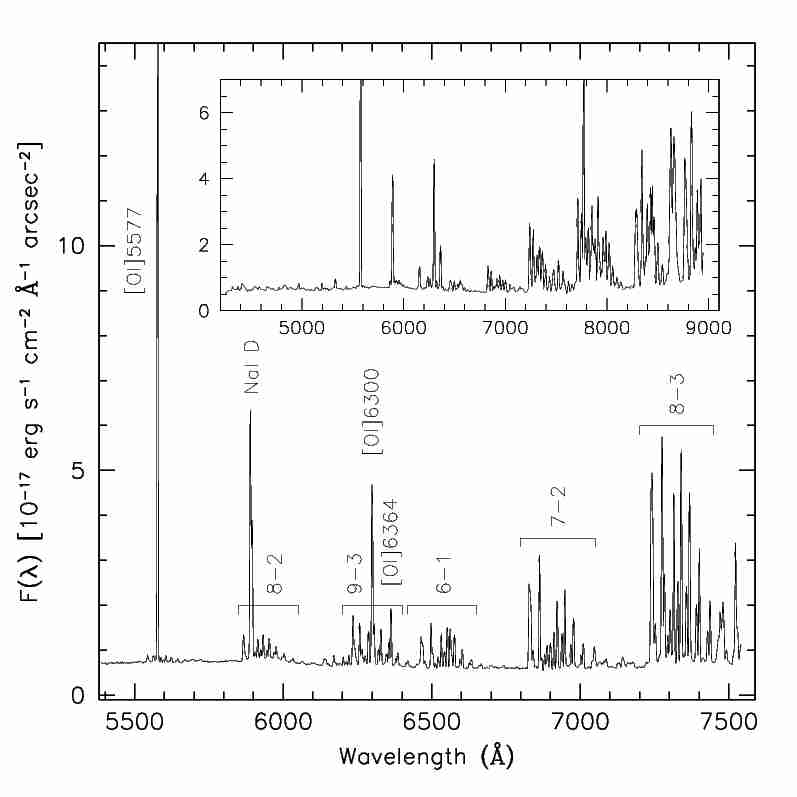}
\caption{\label{fig:skyspectrum}Example of FORS1 flux calibrated night sky 
spectrum obtained with the procedure outlined in the text, for the
600R grism.  Main lines and OH bands identifications are given. The
insert shows an example spectrum obtained with the 300V grism.}
\end{figure}

\subsubsection{\label{sec:extraction}Nightglow spectrum extraction and 
calibration}

After applying the appropriate 2D wavelength solution to all frames,
the night sky spectrum is extracted. For this purpose I have used a
robust algorithm to estimate the mode intensity in each column
perpendicular to the dispersion direction. This implicitly assumes
that most of the pixels are not ``contaminated'' by the contribution
of astrophysical objects, which is reasonable in the majority of the
cases, as verified by direct inspection of the whole two-dimensional data
sample. This is both a consequence of the relatively large
slit length featured by FORS1 (6\am8 on 2048 px) and the typical
targets observed with this instrument, which are very often faint and
star-like sources. After visual inspection, only a few spectra were
removed from the final data set.

To allow for completely unsupervised line and continuum flux
measurements, the accuracy of wavelength calibration is a mandatory
requirement. Possible causes of rigid shifts in the dispersion
solution can be identified as instrument interventions, turning into
movements of the long slit on the focal plane, and flexures at large
zenith distances. To correct for these problems, I have produced a
reference night sky spectrum for each of the two resolutions I have
used, with a typical accuracy (estimated on isolated lines) better
than 1 \AA. Then, by means of cross-correlation, the zero point of the
wavelength scale of each spectrum is automatically corrected at the
end of the extraction procedure. This ensures that, at this stage, all
spectra have maximum wavelength errors that do not exceed 1\AA.

For the absolute flux calibration I have used a set of
spectrophotometric standard stars to derive a reference sensitivity
function $s(\lambda)$, which I have applied to all spectra. Even
though this does not take into account the changes in sensitivity
which are mainly due to the aging of reflective surfaces (Patat
\cite{paperI}), at the wavelengths of interest they are of the order of
a few percent, and therefore can be safely neglected in this context.

The flux calibration of the extracted spectrum $f(\lambda)$ to
physical units is finally computed as:

\begin{displaymath}
F(\lambda) = \frac{f(\lambda)}{s(\lambda) \; t \; w \; p}\;
\;\mbox{erg} \; \mbox{s}^{-1} \;\mbox{cm}^{-2} 
\;\mbox{\AA}^{-1} \;\mbox{arcsec}^{-2}
\end{displaymath}

where $t$ is the exposure time (in seconds), $w$ is the slit width (in
arcsec) and $p$ is the projected pixel scale (arcsec px$^{-1}$).

An example of flux calibrated spectra obtained with this procedure is
presented in Fig.~\ref{fig:skyspectrum}. The resulting signal-to-noise
ratio on the pseudo-continuum changes according to the exposure time
of the original frame, but it is always larger than 100.

\subsubsection{\label{sec:meas}Line/bands flux measurements}

The line flux integration is carried out in a fully automated way
within a given wavelength window after subtracting the estimated
pseudo-continuum intensity. The integration boundaries have been set
according to the spectral resolution. For single lines, like the
[OI]5577, the optimal window semi-amplitude has been set to
1.7$\times$FWHM which corresponds to a $\pm$4$\sigma$ from the line
center. The pseudo-continuum level is estimated in the adjacent
emission line-free regions. For more complex features (Na~ID doublet,
OH bands), the integration boundaries and the continuum region/s have
been set {\em ad hoc}, and are reported in Table~\ref{tab:flux}.
%An example of line integration is shown in Fig.~\ref{fig:fluxmeas}
%for the case of the Na~I D lines.

\begin{figure}
\centering
\includegraphics[width=9cm]{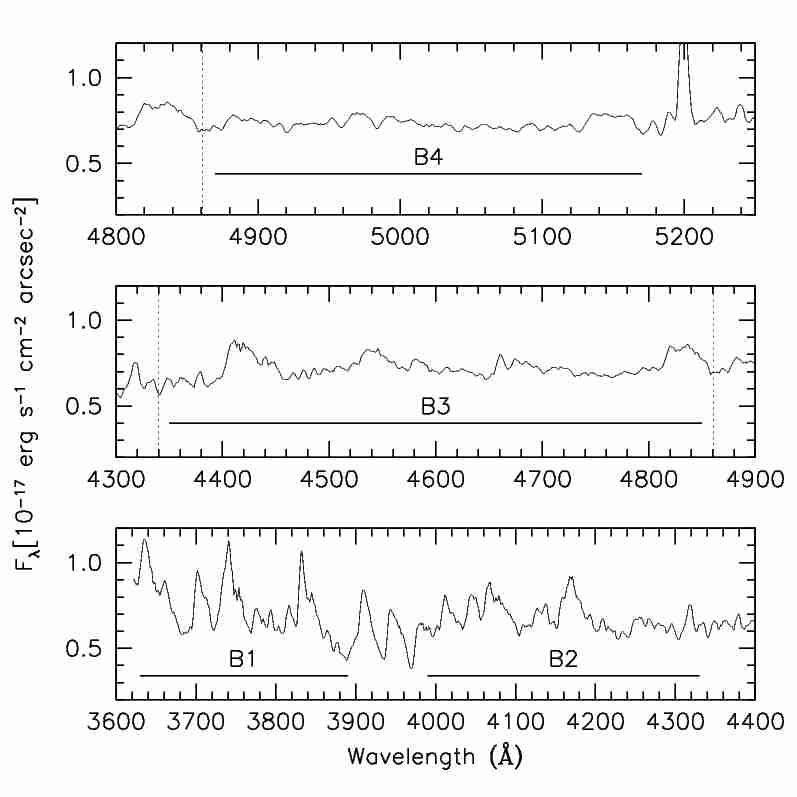}
\caption{\label{fig:bluband}Definition of the blue bands B1, B2, B3 and B4.}
\end{figure}

Following Barbier (\cite{barbier}), I have also introduced four bands,
indicated as B1, B2, B3 and B4, in the blue spectral domain (see
Table~\ref{tab:flux}). The gap between bands B1 and B2 has been set in
order to avoid the strong Ca~II H\&K absorptions
(Fig.~\ref{fig:bluband}).  Finally, a broad band ranging from
5500\AA\/ to 7530\AA\/ has been introduced, with the main aim of
giving an overall flux estimate.

\begin{table}
\caption{\label{tab:flux}Integration boundaries and continuum regions used
for line flux measurements.}
\tabcolsep 1.7mm
\centerline{
\begin{tabular}{ccccc}
\hline \hline
Line/Band   & Line Range & Cont. Range & & Setup \\
\hline
NI  5200    & $\pm 4\sigma$ & 5040-5120 &(C1)   & 300V,600B\\
OI] 5577    & $\pm 4\sigma$ & 5480-5520 &(C2)   & 300V,600B/R\\ 
Na~I~D       & D$_2-4\sigma$-D$_1+4\sigma$ & 5800-5850 &(C3) & 300V,600B/R\\
OI] 6300    & $\pm 4\sigma$ & 6400-6450 &(C4)   & 300V,600R\\ 
OI] 6364    & $\pm 4\sigma$ & 6400-6450 &       & 300V,600R\\ 
\hline
OH(6-1)     & 6435-6680     & 6750-6800 &(C5)   & 300V,600R\\
OH(7-2)     & 6810-7060     & 6750-6800 &       & 300V,600R\\
OH(8-3)     & 7200-7450     & 6750-6800 &       & 300V,600R\\
OH(6-2)     & 8250-8570     & 8160-8230 &(C6)   & 300V+GG\\
O$_2$(0-1)  & 8605-8695     & 8160-8230 &       & 300V+GG\\
\hline
B1          & 3630-3890     & -         & & 300V,600B\\
B2          & 3990-4330     & -         & & 300V,600B\\
B3          & 4350-4850     & -         & & 300V,600B\\
B4          & 4870-5170     & -         & & 300V,600B\\
\hline
Broad Band  & 5500-7530     & -         & & 300V,600R \\
\hline
\end{tabular}
}
\end{table}

As in the case of the broad-band measurements, the line fluxes need to be
corrected for airmass. Between the two cases, however, there is a
difference: in fact, while the integrated flux within a broad-band
filter is the result of extra-terrestrial sources (zodiacal light,
unresolved stars and galaxies) and emission within the atmosphere, in
the case of nightglow emission lines all the radiation is of
terrestrial origin. Practically this coincides with setting $f=1$ in
Eq.~C.3 of Paper~I, which I have used to correct the observed values.

In general, the signal in the measured features is so high that the
uncertainty in the line fluxes is by far dominated by the
contamination by unresolved OH lines and, to a smaller extent, to the
uncertainty on the pseudo-continuum level.

\section{\label{sec:dark}Dark time night sky brightness}

\begin{figure}
\centering
\includegraphics[width=9cm]{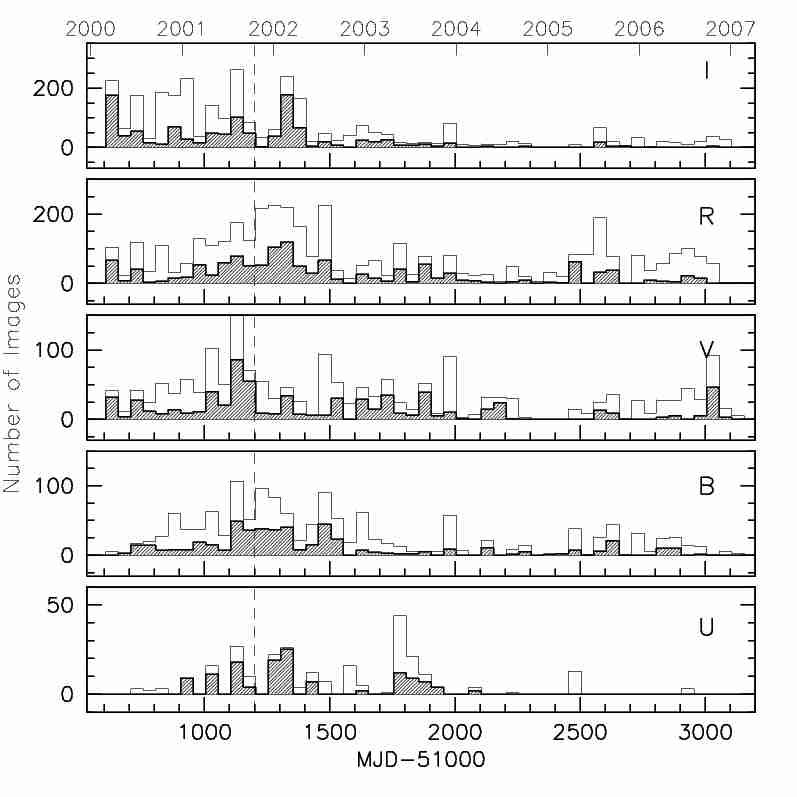}
\caption{\label{fig:hist} Image distribution along the time interval 
covered by the present work for the global data set (thin line) and
for the dark-time set (shaded thick line). The vertical dotted line
marks the extent of the sample presented in Paper I, while the upper
scale marks the 1st of January of each year.}
\end{figure}

Since the data set includes observations obtained under a wide variety
of conditions, in order to estimate the zenith sky brightness during
dark time it is necessary to apply some filtering. To this aim I have
adopted the same criteria described in Paper I: photometric
conditions, airmass $X\leq$1.4, galactic latitude $|b|>$10$^\circ$,
helio-eclitpic longitude $|\lambda-\lambda_\odot|\geq$90$^\circ$, time
distance from the closest twilight $\Delta t_{twi}>$ 1 hour and no
moon (fractional lunar illumination FLI=0 or moon elevation
$h_M\leq-$18$^\circ$). The results of this selection, which reduced
the number of suitable data points to 3736, are summarized in
Table~\ref{tab:dark}. As one can see, the average values are all
within 0.1 mag from those reported in Paper I (see Table~4). In all
filters there is a systematic shift towards darker values, with the
only exception of the $I$ band. Since the values reported in Paper I
were obtained during the sunspot maximum and, given the correlation
between solar activity and night sky brightness shown by Walker
(\cite{walker88}), Pilachowski et al. (\cite{pila}); Krisciunas
(\cite{krisc90}), Leinert et al. (\cite{leinert95}) and Mattila et
al. (\cite{attila}, Krisciunas (\cite{krisc97}), Krisciunas et
al. (\cite{krisc07}), this behavior was indeed expected. Given the time
distribution of data points (see Fig.~\ref{fig:hist}), the average
values reported in Table~\ref{tab:dark} are biased towards the sunspot
maximum phase.  For solar activity corrected data, see next section.

\begin{table}
\centering
\caption{\label{tab:dark} Zenith corrected average sky brightness 
during dark time at Paranal. Values are expressed in mag
arcsec$^{-2}$.  Columns 3 to 8 show the RMS deviation, minimum and
maximum brightness, number of dark-time data points, expected average
contribution from the zodiacal light, and total number of data points,
respectively.}
\begin{tabular}{cccccccccc}
\hline \hline
Filter & Sky Br. & $\sigma$ & Min & Max & $N_{dark}$ & $\Delta m_{ZL}$ & $N_{tot}$\\
\hline
U & 22.35 & 0.19 & 21.89 & 22.78  &  129 &0.20 & 264\\
B & 22.67 & 0.16 & 22.19 & 23.02  &  493 &0.28 & 1400\\ 
V & 21.71 & 0.24 & 21.02 & 22.30  &  692 &0.20 & 1836\\
R & 20.93 & 0.24 & 20.42 & 21.56  & 1285 &0.16 & 3931\\
I & 19.65 & 0.28 & 18.85 & 20.56  & 1137 &0.07 & 3001\\
\hline 
Total &   &      &       &        & 3736 &     & 10432\\
\hline
\end{tabular}
\end{table}

\begin{figure}
\centering
\includegraphics[width=9cm]{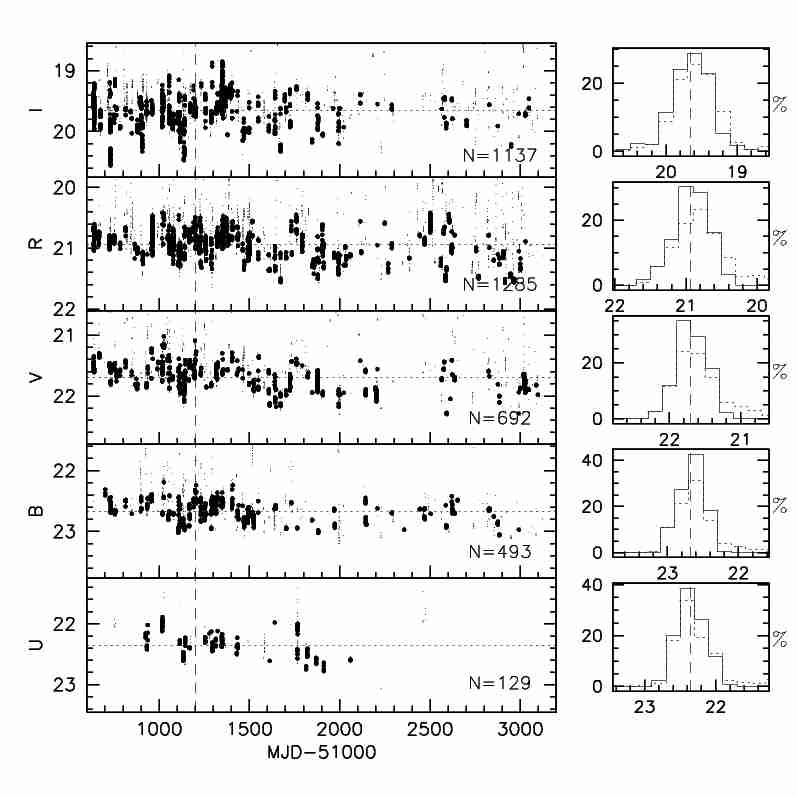}
\caption{\label{fig:dark} Zenith corrected sky brightness measured at 
Paranal during dark time (thick dots) from April 1st, 2000 to April 8,
2006.  The selection criteria are: $|b|>$10$^\circ$,
$|\lambda-\lambda_\odot|\geq$90$^\circ$, $\Delta t_{twi}>$1 hour,
FLI=0 or $h_m\leq-$18$^\circ$.  Thin dots indicate all observations
(corrected to zenith). The horizontal dotted lines are positioned at
the average values of the selected points, while the vertical dashed
line marks the end of the time interval discussed in Paper I.  The
histograms trace the distribution of selected measurements (solid
line) and all measurements (dotted line), while the vertical dashed
lines are placed at the average sky brightness during dark time.}
\end{figure}

Single measurements for the 5 passbands are presented in
Fig.~\ref{fig:dark}, which shows also surface brightness distributions
for the dark time (solid line) and global (dashed line) samples.

\begin{figure}
\centering
\includegraphics[width=9cm]{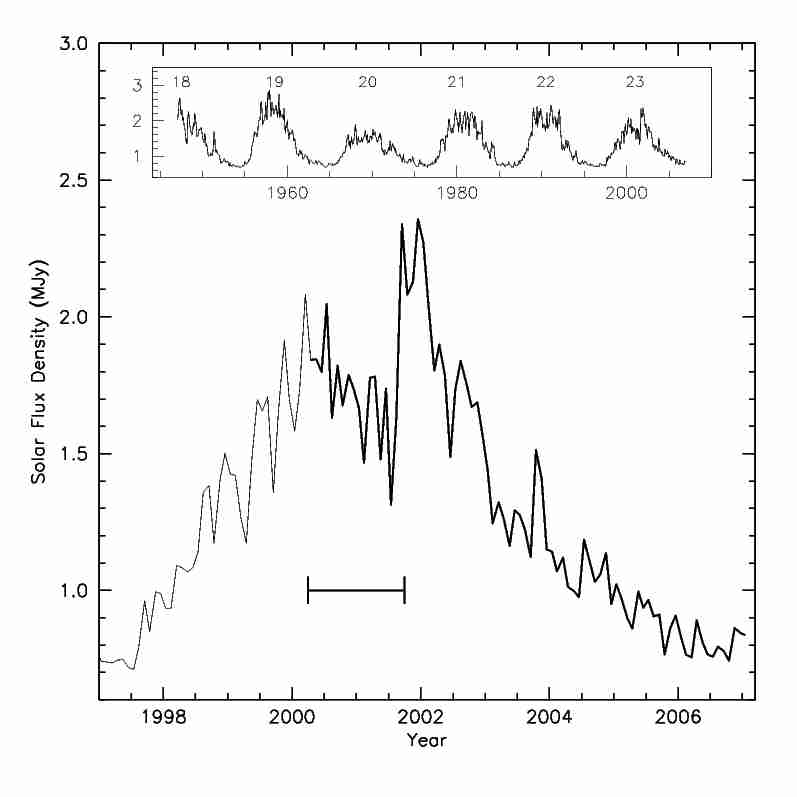}
\caption{\label{fig:sun2}Penticton-Ottawa Solar flux at 2800 MHz 
(monthly average). The time range covered by the data presented in
this paper is indicated by the thick line, while the horizontal
segment indicates the time covered by the data presented in Paper
I. The upper insert traces the solar flux during the last six cycles.}
\end{figure}

\begin{figure}
\centering
\includegraphics[width=9cm]{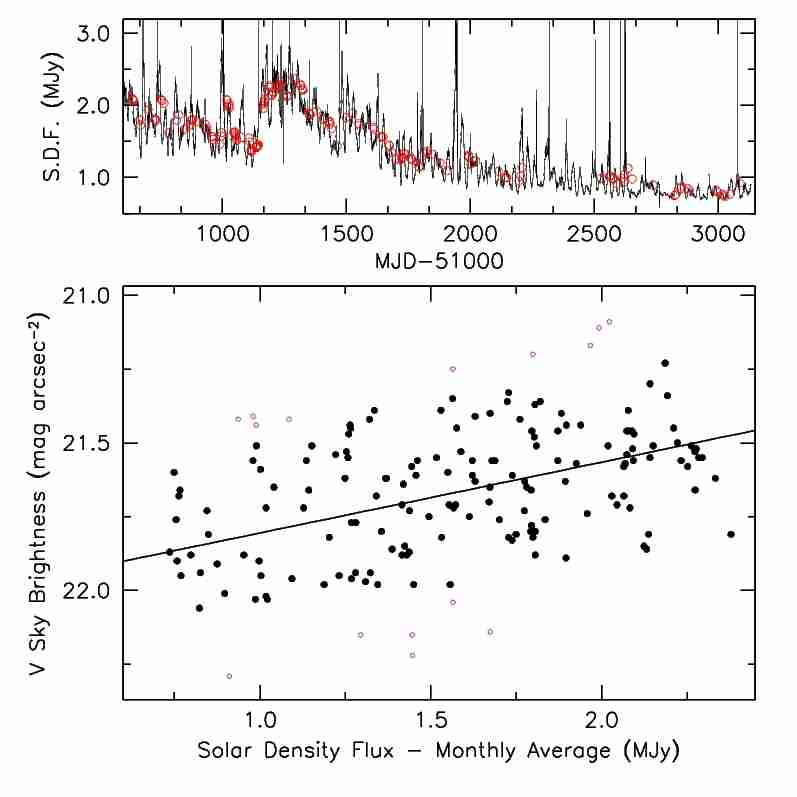}
\caption{\label{fig:sunave}Lower panel: nightly average dark time sky 
brightness in the $V$ passband vs. solar flux density. The solid line
is a linear least squares fit to the data (empty symbols mark the data
points rejected by a kappa-sigma clipping). Upper panel:
Penticton-Ottawa solar flux at 2800 MHz during the time interval
discussed in this paper. The open circles indicate the monthly
averaged values corresponding to the $V$ nightly averages plotted in
the lower panel.}
\end{figure}

\section{\label{sec:sun}Sky brightness vs. solar activity}

As first pointed out by Rayleigh (\cite{rayleigh}) and confirmed later
on by several other authors (see for instance Rosenberg \& Zimmerman
\cite{rosenberg}; Walker \cite{walker88}; Krisciunas \cite{krisc90}; 
Leinert et al. \cite{leinert95}; Mattila et al. \cite{attila};
Krisciunas \cite{krisc97}; Krisciunas et al. \cite{krisc07}) many of
the emission features in the night sky spectrum show a clear
dependency on the sunspot cycle. In particular, $B$ and $V$ present a
peak-to-peak variation of $\sim$0.5 mag arcsec$^{-2}$ during a full
solar cycle. Less clear is the behavior at longer wavelengths, which
are dominated by the OH emissions, whose intensity is uncorrelated
with solar activity (see Sec.~\ref{sec:spectra}).

The data presented here cover the descent from the maximum of sunspot
cycle n.~23 to the minimum phase, as shown in Fig.~\ref{fig:sun2},
that displays the monthly averaged Penticton-Ottawa solar flux at
2800 MHz (Covington 1969)\footnote{The data are available in digital
form at the following web site: {\tt
http://www.drao.nrc.ca/icarus/www/archive.html}}. During this interval
the Solar Flux Density (hereafter $SFD$) spans from 0.8 to 2.4 MJy, a
range which is very close to that of a full cycle (the solar minimum
is expected for the end of 2007). Following what has been done by
other authors (see for instance Leinert et al. \cite{leinert95}), I
have studied the correlation between the sky brightness nightly
averages and the SFD monthly averages, computed during the 30
preceding days. The sky brightness measurements have been corrected
for the zodiacal light contribution computed for each data point as in
Paper~I (Sec.~4) and using the data by Levasseur-Regourd \& Dumont
(\cite{levasseur}).

All passbands show very good linear correlations, an example of which
is presented in Fig.~\ref{fig:sunave} for the $V$ filter. In order to
give a quantitative representation of the effect, I have fitted a
relation of the type $m=m_0 + \gamma \;SFD$ to the data. The results
are shown in Table~\ref{tab:sunave} for all filters. Besides reporting
the zeropoint ($m_0$, mag arcsec$^{-2}$), the slope ($\gamma$, mag
arcsec$^{-2}$ MJy$^{-1}$) and their associated statistical errors, the
Table includes also the estimated full solar cycle variation ($\Delta
m=|(2.4-0.8)\gamma|$), the value attained at solar minimum, evaluated
for $SFD$=0.8 MJy ($m_{min}$), the value corresponding to the average
$SFD$ level $<SFD>$=1.6 MJy ($m_{ave}$), the RMS deviation from the
best fit relation ($\sigma$), the linear correlation factor ($r$) and
the number of nights used ($N$) for each filter.

As one can see, the values of $\Delta m$ are smaller than those
reported by other authors: with the only exception of $U$, which
reaches about 0.6 mag arcsec$^{-2}$, all the others show values that
are smaller than 0.4 mag arcsec$^{-2}$.  Walker (\cite{walker88})
quoted maximum ranges of $\Delta V\simeq$1.0 and $\Delta B\simeq$0.8
mag arcsec$^{-2}$ for solar cycle n.~21, while Krisciunas
(\cite{krisc97}) reports $\Delta V$=0.6 for solar cycle n.~22, and
similar values are reported by Leinert et al. (\cite{leinert95}) and
Mattila et al. (\cite{attila}).  On the other hand, Liu et
al. (\cite{liu}) quote an increase of the $V$ sky brightness of
$\sim$0.2 mag arcsec$^{-2}$ from 1995 to 2001.  This value is
consistent with the measures discussed here, especially taking into
account that cycle n.~23 had a second maximum, which occurred after
the observations presented by Liu et al. (\cite{liu}).  These facts
seem to suggest that not all solar cycles have identical effects on
the night glow. As a matter of fact, Walker (\cite{walker88}), while
revising the result of previous works, had suggested that the relation
between intensity of the [OI]5577
\AA\/ line and the solar activity might vary from cycle to cycle,
within a given cycle and possibly with geographical location.
Unfortunately, the number of sunspot cycles covered by the
observations is still too small to allow a firm conclusion, but the
very recent results discussed by Krisciunas et al. (\cite{krisc07}),
covering two full solar cycles, seem indeed to confirm this suspicion.

\begin{figure}
\centering
\includegraphics[width=9cm]{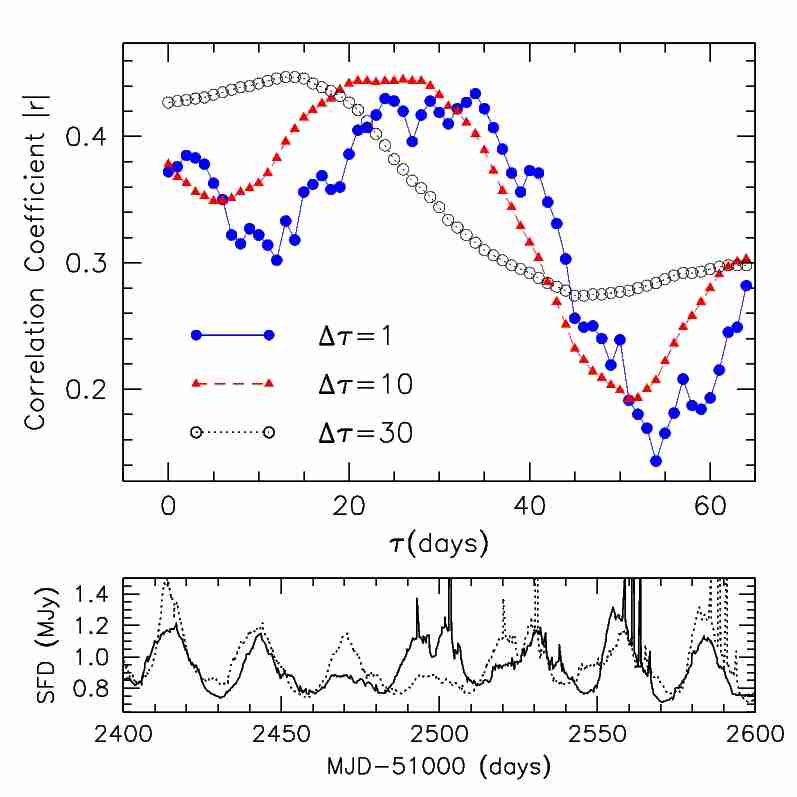}
\caption{\label{fig:sunanal}Upper panel: linear correlation coefficient 
(absolute value) as a function of time delay $\tau$ for the $V$
passband data, computed for three different values of $\Delta \tau$
(1, 10 and 30 days). Lower panel: example of SFD periodicity due to
solar rotation. The dotted curve is a replica of the original data
shifted by 27.3 days.}
\end{figure}

\subsection{\label{sec:timescale} Solar activity correlation time scales}

Given the unprecedentedly large amount of data, one can investigate
the solar dependency in a bit more detail, for example trying to
deduce the typical timescales of night sky brightness fluctuations due
to short-term changes in the solar flux. Ideally, for doing this, one
would look for strong enhancements in the solar flux and try to seek
for a corresponding increases in the night sky
brightness. Nevertheless, due to the sparse time sampling, the only
viable approach is actually the opposite one, i.e. starting from the
available night sky measurements, one goes back to the solar data and
studies the correlation with the sun flux emitted as a function of
time lag. For doing this I have computed the Pearson's linear
correlation coefficient $r$ (Press et al. \cite{press}) between the
nightly average sky brightness $m(t)$ measured at any given time $t$
and the solar flux density $SFD(t-\tau)$, averaged within a time
window $\Delta \tau$, as a function of both $\tau$ and $\Delta
\tau$. While the first parameter gives an indication about the time
lag between a change in the solar flux and the consequent variation in
the night sky brightness, the second is related to the typical
timescales of the physical processes which govern the energy release
in the upper atmosphere.

An example of this kind of analysis is shown in the upper panel of
Fig.~\ref{fig:sunanal}, which illustrates the behavior of the
correlation coefficient $r$ for the $V$ data, obtained on 170 separate
nights for three different values of $\Delta \tau$ (1, 10 and 30
days). The correlation peak is quite broad and it is centered at
$\tau\sim$30 days; moreover, the maximum correlation tends to increase
slightly for larger averaging windows. This plot also explains why the
average solar flux computed in the 30 days before the night sky
observations (empty circles) gives a better correlation than the sun
flux measured on the preceding day (filled circles), as already
pointed out by several authors (see for example Leinert et
al. \cite{leinert95}; Mattila et al. \cite{attila}). Finally,
Fig.~\ref{fig:sunanal} clearly shows that the correlation drops
significantly for $\tau>$40 days, a fact that is common to all
$UBVRI$ passbands.

An interesting feature to be noticed in Fig.~\ref{fig:sunanal} (seen
also in the equivalent plots for the other filters), is the presence
of spurious correlation peaks at a constant separation of about 27
days from the main peak. This is due the a periodicity present in the
solar flux data (see Fig.~\ref{fig:sunanal}, lower panel), which is
related to the solar rotation, whose synodic period is $\sim$27.3 days
(Howard
\cite{howard}). This semi-regular recurrence in the solar data explains,
for example, the presence of the two bumps close to $\tau$=5 and
$\tau$=65 in Fig.~\ref{fig:sunanal}.

The strongest correlation is shown by the $U$ passband data, which
presents a rather marked peak $r\sim$0.6 at $\tau\sim$15 days, while
$r\sim$0.15 for $\tau$=1 day. For this reason, averaging over the last
30 days gives a strong increase in the correlation, much stronger than
in any other passband. A behaviour similar to that displayed in $V$ is
seen also in $B$ (correlation peak $r\sim$0.45 for $\tau\sim$25 days)
and $I$ ($r\sim$0.30 for $\tau\sim$20 days). Somewhat different is the
case of $R$ passband, for which the correlation peak ($r\sim$0.45) is
attained at $\tau\sim$2 days, suggesting that the sun-dependent
features that contribute to the flux in this filter react rather
rapidly to the solar flux fluctuations.  In general, however, and with
the possible exception of the $U$ passband, the correlation peaks are
rather broad, indicating that different processes take place with
different timescales. The behavior of the $U$ band, where the
nightglow emission is dominated by the Herzberg and Chamberlain O$_2$
bands (Broadfoot \& Kendall
\cite{broadfoot}), indicates that the photo-chemical reactions that 
are responsible for the emission in this region are more sensitive to
solar activity.

\begin{table}
\tabcolsep 0.7mm
\caption{\label{tab:sunave} Linear least squares fit parameters for the sky 
brightness vs.solar activity relation. Values in parenthesis indicate
RMS uncertainties. Input data have been corrected for the differential
zodiacal light contribution (see text).}
\centering
\begin{tabular}{ccccccccc}
\hline \hline
Filter & $m_{min}$ &$m_{ave}$& $\Delta m$ & $m_0$ & $\gamma$ & $\sigma$ & $r$ & N \\
\hline
%U      & 22.63 & 22.40 & 0.50 & 22.87 (0.10)& $-$0.29 (0.05)& 0.12 & 32  \\
%B      & 22.81 & 22.68 & 0.29 & 22.95 (0.04)& $-$0.17 (0.02)& 0.12 & 123 \\
%V      & 21.82 & 21.66 & 0.34 & 21.98 (0.05)& $-$0.20 (0.03)& 0.17 & 154 \\
%R      & 21.14 & 20.96 & 0.40 & 21.33 (0.04)& $-$0.24 (0.03)& 0.17 & 202 \\
%I      & 19.76 & 19.65 & 0.24 & 19.88 (0.06)& $-$0.14 (0.04)& 0.19 & 145 \\
U      & 22.86 & 22.58 & 0.61 & 23.15 (0.12)& $-$0.36 (0.07)& 0.15 & 0.47 & 32  \\
B      & 23.11 & 22.98 & 0.29 & 23.25 (0.04)& $-$0.17 (0.02)& 0.12 & 0.40 & 127 \\
V      & 21.99 & 21.86 & 0.30 & 22.13 (0.05)& $-$0.17 (0.03)& 0.14 & 0.42 & 148 \\
R      & 21.26 & 21.09 & 0.37 & 21.33 (0.04)& $-$0.22 (0.02)& 0.15 & 0.44 & 202 \\
I      & 19.81 & 19.72 & 0.20 & 19.90 (0.06)& $-$0.11 (0.04)& 0.18 & 0.28 & 144 \\
\hline
\end{tabular}
\end{table}

\section{\label{sec:season}Seasonal Variations}

\begin{figure}
\centering
\includegraphics[width=9cm]{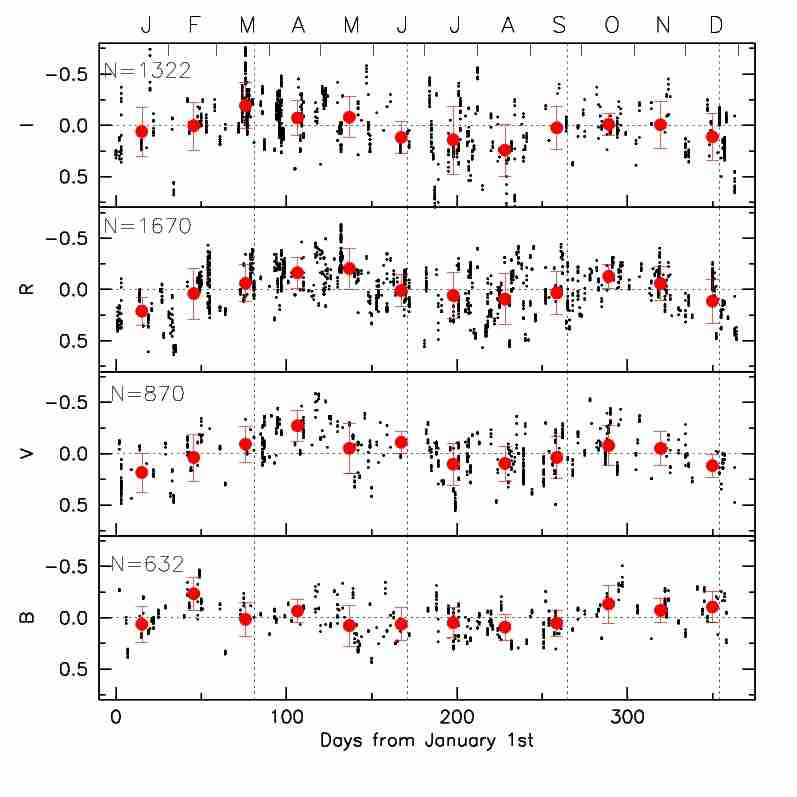}
\caption{\label{fig:season}Seasonal variation of dark time sky
brightness with respect to the average value. Small symbols are the
single measurements while the large symbols mark the monthly averages,
computed in bins of 30 days each. The error bars indicate the RMS
deviations from the average within each bin, while the vertical dotted
lines mark equinoxes and solstices. Data have been corrected for
differential zodiacal light contribution and solar flux dependency.}
\end{figure}

In the previous work I had attempted to detect night sky brightness
seasonal variations but, due to insufficient number of data points, I
could not draw any firm conclusion (see Fig.~14 in Paper I). Thanks to
the much larger sample now available, this analysis becomes feasible
and, as a matter of fact, traces of a periodic modulation in the
average sky brightness are visible already in
Fig.~\ref{fig:dark}. They become much clearer when each data point is
plotted against the number of days from the beginning of the
corresponding year. The result is shown in Fig.~\ref{fig:season}
where, besides reporting the single dark time measurements, I have
also plotted the monthly averages. The input data have been corrected
for differential zodiacal light contribution and the solar flux
dependency derived in the previous section has been removed using the
parameters presented in Table~\ref{tab:sunave}. 

This semi-annual oscillation (hereafter SAO) is definitely present in
$V$, $R$ and $I$, while its presence in $B$ is more questionable ($U$
data were not included since the sample in this passband is too poor
for this purpose). The modulation amplitude grows at longer
wavelengths, shows two maxima around April-May and October, and two
minima around July-August and December-January. In general, the
variation is more pronounced in Winter-Spring than in Summer-Fall. For
example, in $I$ it reaches a peak-to-peak value of about 0.5 mag
arcsec$^{-2}$.

As pointed out by Benn \& Ellison (\cite{benn}), the variable
contribution of zodiacal light can mimic a seasonal variation. In
order to exclude a possible contribution by this source to the
observed behavior, I have analyzed the expected enhancement of
brightness due to the zodiacal light for each data point, using the
data presented by Levasseur-Regourd \& Dumont (\cite{levasseur}) and
the procedure discussed in Paper I (Sec.~4) for the Paranal
site. This does not show any significant regular pattern as a function
of day of the year. The conclusion is that the observed SAO is not due
to the periodic apparent variation of the ecliptic height above
Paranal's horizon.

An interesting thing to be noticed, is that the minima and maxima of
the SAO occur out of phase with respect to the Equinoxes and Solstices
(see Fig.~\ref{fig:season}, vertical dotted lines).

While seasonal variations of emission lines and/or bands have been
studied by several authors in the past (see for example Chamberlain
\cite{chamberlain}, Roach \& Gordon \cite{roach} and references
therein), broad band measurements are much more scanty and the results
not always in agreement. For instance, Schneeberger, Worden \& Beckers
(\cite{schneeberger}) report particularly bright values obtained in
June at the Sacramento Peak Observatory and they find them to be {\em
marginally correlated with the strong seasonal trend evident in the
record of daytime sky brightness observations}. In their survey run at
the Lowell Observatory, Lockwood, Floyd \& Thompson (\cite{lockwood})
discuss the seasonal variation, concluding that {\em neither winter
enhancements [...] nor springtime rise [...] is indicated [...]}. Benn
\& Ellison (\cite{benn}) reach the same result from the analysis of
the data obtained on La Palma, concluding that {\em dark-of-moon sky
brightness does not vary significantly ($\leq$0.1 mag) with season
[...]}.  Finally, Liu et al. (\cite{liu}), analyzing data taken at the
Xinglong Station between 1995 and 2001, find that {\em the sky is
darker in the fall and winter than in the spring and summer
[...]}. While part of the discrepancies can be due to latitude effects
(see for example Chamberlain \cite{chamberlain}), some of the negative
detections are probably to be ascribed to non sufficient time sampling
and coverage. In fact, the SAO amplitude is at most comparable with
the night-to-night fluctuations and hence large and well sampled data
sets are required.

It is worth mentioning here that Garstang (\cite{garstang88}) has
produced some simplified models to predict seasonal variations in the
broad-band night sky brightness, based on periodic variations of
height and molecular density. The values predicted by these models are
far too small with respect to those presented here, suggesting that
other possible explanations must be investigated.

The observed behavior might indicate that whatever the reason for the
periodic variation is, it is not directly related to the amount of sun
radiation received by a given patch of the atmosphere during the day.
In fact, one might think that since during the austral summer days are
much longer than nights, this could result into a brighter nightglow.
The data show actually the opposite behavior, since during austral
summer the night sky reaches its lowest average brightness. Moreover,
this appears to be in phase with what is observed in the northern
hemisphere, where the sky is darker in winter than is summer (see for
example Liu et al. \cite{liu}). This seems to indicate that the SAO
must be related to some other, non local mechanism (see the discussion
in Sec.~\ref{sec:disc}). Remarkably, but after all not surprisingly,
polar auroral activity shows a similar temporal fluctuation, with
maxima in spring and autumn (see for example Meinel, Neighed
\& Chamberlain \cite{meinel}).

\section{\label{sec:spectra}Spectroscopic analysis}

Since there are many distinct components that contribute to the global
nightglow emission (Leinert et al. \cite{leinert} and references
therein) that cannot be disentangled with broad-band photometry,
several attempts have been made in the past to increase the spectral
resolution using a set of narrow band filters.  After the pioneering
work by Barbier (\cite{barbier}; see also Chamberlain
\cite{chamberlain} for a review on this subject), who used 8
intermediate-band photometry, several other researchers have
undertaken similar projects (see for example Leinert et
al. \cite{leinert95} and Mattila et al. \cite{attila} for two more
recent works). Nevertheless, due to the relatively large bandwidths
(typically broader than 100 \AA), the study of single features,
especially if not very intense, has always been hindered by the
presence of the pseudo-continuum and possible adjacent lines and/or
bands.

\begin{figure}
\centering
\includegraphics[width=9cm]{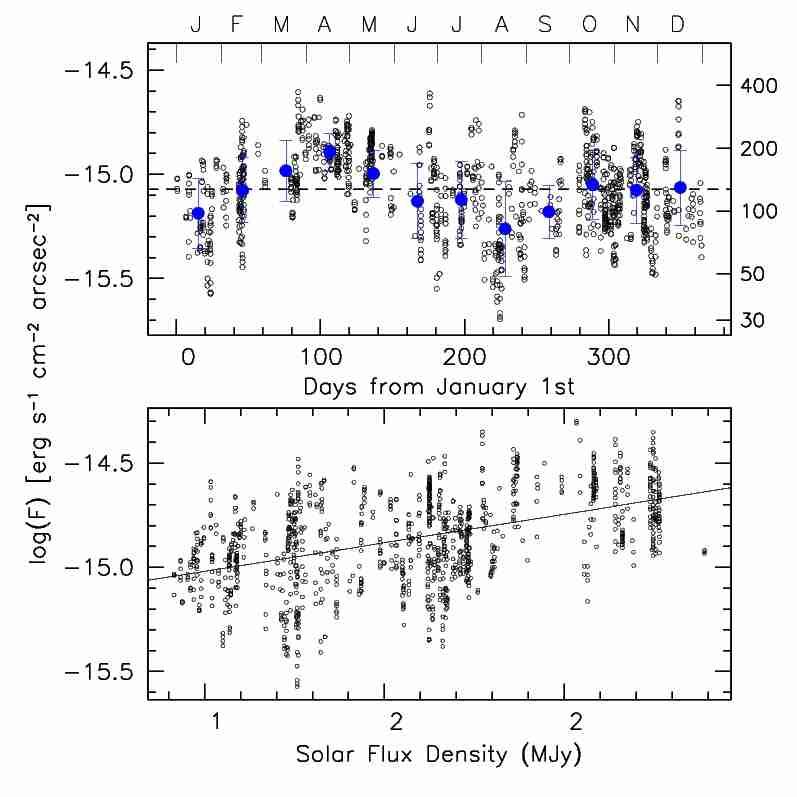}
\caption{\label{fig:oi5577}Lower panel: dark time, zenith corrected 
[OI]5577 line flux as a function of $SFD$. The solid line traces
a linear least-squares fit to the data. Upper panel: line fluxes as a
function of time from the beginning of the year. The data have been
corrected to solar minimum ($SFD$=0.8 MJy) using the relation shown in
the lower panel. The large points mark monthly averages and the right
vertical scale is expressed in Rayleigh.}
\end{figure}

\subsection{\label{sec:lines}Main atomic emission lines}

In this section I will analyze the behavior of the main emission lines
in the optical domain. For convenience, I will express the line fluxes
in Rayleighs\footnote{1R=10$^6/4\pi$ photons s$^{-1}$ cm$^{-2}$
sr$^{-1}$ $\equiv$ 3.72$\times$10$^{-14}\; \lambda^{-1}$(\AA) erg
s$^{-1}$ cm$^{-2}$ arcsec$^{-2}$.}.

\subsubsection{\label{sec:oi5577}The [OI] lines}

The [OI]5577 is generally the most prominent feature in the
optical night sky spectrum. It falls right in the center of the $V$
passband, giving a typical contribution of 20\% to the global surface
brightness in this filter. It has a typical intensity of $\sim$250 R,
it arises in layers placed at about 90 km (Roach \& Gordon
\cite{roach}) and it displays a marked dependency on solar activity
(Rayleigh \cite{rayleigh}). This is clearly shown also by the data
presented here (Fig.~\ref{fig:oi5577}, lower panel), which indicate
also the presence of pronounced fluctuations (40 to 750 R
peak-to-peak) around the average level ($\sim$230 R). To quantify the
correlation with solar activity and following the procedure that has
been applied to the broad band data (see Sec.~\ref{sec:sun}), I have
fitted to the data a law of the type $\log F = \log F_0 + \gamma \;
SFD$. The results for this and other lines are presented in
Table~\ref{tab:lines}, which shows also the number of data points used
($N$) and the linear correlation coefficient ($r$). What is
interesting to note is that, once the solar dependency is removed from
the data, the flux of this line displays a marked SAO (see
Fig.~\ref{fig:oi5577}, upper panel), similar to that seen in the broad
band data (Fig.~\ref{fig:season}). This fact has been already noticed
by Buriti et al. (\cite{buriti}), who found that this line and other
mesospheric features all show a SAO.

\begin{figure}
\centering
\includegraphics[width=9cm]{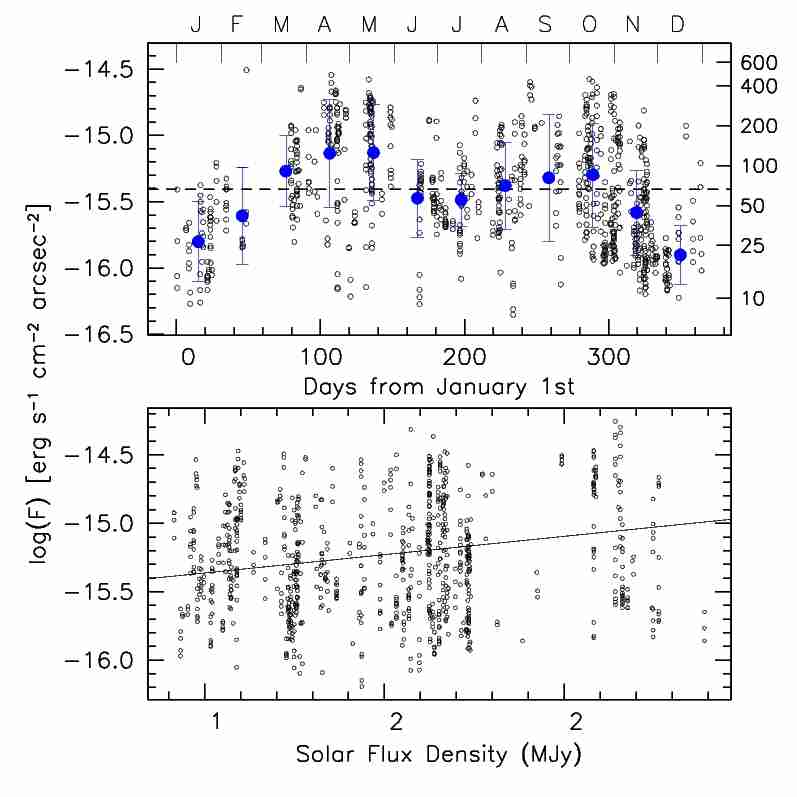}
\caption{\label{fig:oi6300}Same as Fig.~\ref{fig:oi5577} for the 
[OI]6300 line.}
\end{figure}

\begin{figure}
\centering
\includegraphics[width=9cm]{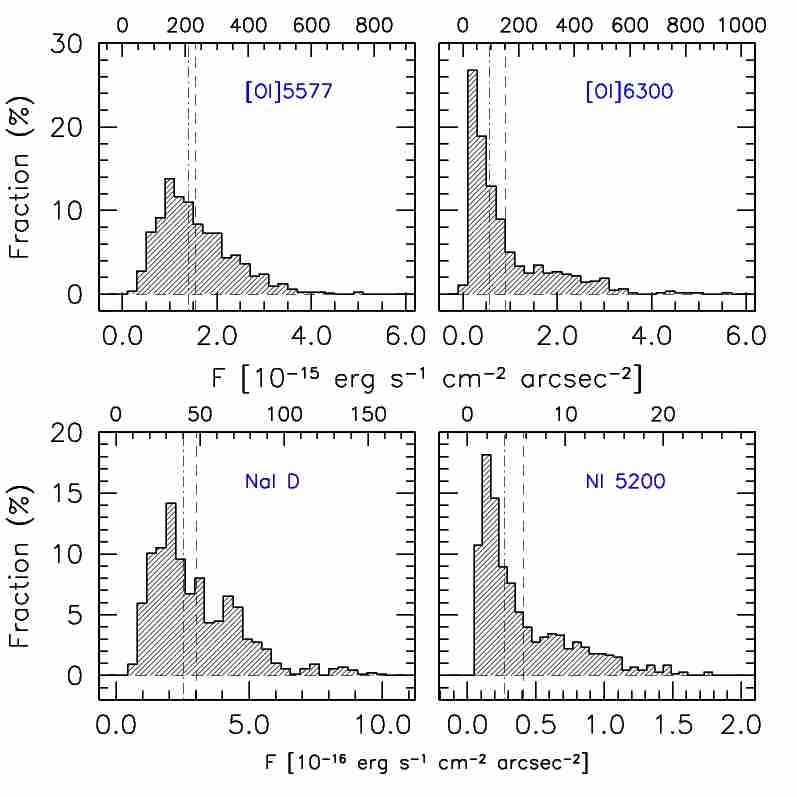}
\caption{\label{fig:linehist}Dark time line flux distributions for 
[OI]5577\, [OI]6300,6364\AA, Na~I~D and N~I 5200. The vertical
lines trace the average (dashed) and the median (dotted-dashed) of the
distributions. The upper horizontal scale in each panel is expressed
in Rayleigh.}
\end{figure}

The same behavior is, in fact, shown by the [OI]6300,6364\AA\/ doublet (see
Fig.~\ref{fig:oi6300}), which is produced at 250-300 km (Roach \&
Gordon \cite{roach}) and it is known to undergo abrupt intensity
changes on two active regions about 20$^\circ$ on either side of the
geomagnetic equator (Barbier \cite{barbier2}), hence marginally
including the Paranal site. Indeed, the [OI]6300 measured fluxes
show very strong variations (10 to 950 R) around the average level
($\sim$150 R), with spikes reaching $\sim$1 kR, so that the line
fluxes span almost a factor 100 (to be compared with the factor
$\sim$25 measured for the [OI]5577. In this respect, the behavior
of the [OI]6300,6364\AA\/ doublet is different from that of the
[OI]5577 line, since in a significant number of cases its flux is
very small and the line is practically invisible, lost in the OH(9-3)
molecular band (see Fig.~\ref{fig:skyspectrum}). In fact, its flux is
less than 80 R for more than 50\% of the cases considered in this
work, the minimum recorded value being $\sim$10 R
(Fig.~\ref{fig:linehist}).  The variation range of these two features
and their relative contributions to the $V$ and $R$ total flux are
summarized in Table~\ref{tab:oi}. From the values reported in this
table, one can see that [OI]5577 can produce a maximum variation
of $\sim$ 0.5 mag arcsec$^{-2}$ in the $V$ band, while a similar
effect is produced by the [OI] doublet in the $R$
passband\footnote{Since the [OI]6364/[OI]6300 lines ratio is 1/3,
the maximum combined contribution is about 60\%.}. Given the fact that
the RMS variation in the $V$ and $R$ passbands is about 0.25 mag
arcsec$^{-2}$ (see Table
\ref{tab:dark}), this implies that the fluctuations seen in these passband
are not completely accounted by the changes in the atomic O line
fluxes.  The two lines appear to show a very weak correlation: the
average ratio F(6300)/F(5577) is 0.64, but in a significant number of
cases ($\sim$25\%) this ratio is larger than 1 (see also
Sec.~\ref{sec:corr}). In those circumstances, the [OI]6364\AA\/ line
becomes the most prominent nightglow atomic feature in the optical
domain.

\begin{table}
\tabcolsep 2mm
\caption{\label{tab:oi}Observed line intensities and relative flux 
contribution to $V$ and $R$ passbands for [OI]5577, [OI]6300
and Na~I D. All values are corrected to zenith.}
\begin{tabular}{ccccc}
\hline
\hline
   & \multicolumn{2}{c}{F}                      & V  & R \\
{\bf [OI]5577}  & (10$^{-16}$ erg s$^{-1}$ cm$^{-2}$ arcsec$^{-2})$ & ($R$) &    & \\
\hline
min  & 2.6  & 40  & 3\%  & $<$1\%\\
ave  & 16   & 230 & 20\% & 1\% \\
max  & 50   & 750 & 62\% & 4\% \\
     &      &     &      &     \\
\hline
{\bf [OI]6300} & & & & \\
min  & 0.6  & 10  & $<$1\%  & $<$1\%\\
ave  & 9    & 150 & 2\%    & 7\% \\
max  & 56   & 950 & 12\%   & 45\% \\
     &      &     &      &     \\
\hline
{\bf Na~I D} & & & & \\
min  & 0.7  & 10  & $<$1\%  & $<$1\%\\
ave  & 3    & 50  & 2\%     & 3\% \\
max  & 10   & 160 & 8\%     & 9\% \\
\hline
\end{tabular}
\end{table}

As far as the solar activity is concerned, it must be noticed that
even though a dependence from the $SFD$ is seen, it is less clear than
in the case of the [OI]5577 (see Fig.~\ref{fig:oi6300}, lower
panel). In fact, even though the slopes are very similar, the linear
correlation coefficient for [OI]6300 is 0.23, to be compared with
0.49 measured for the [OI]5577 line (see also
Tab.~\ref{tab:lines}).

\begin{figure}
\centering
\includegraphics[width=9cm]{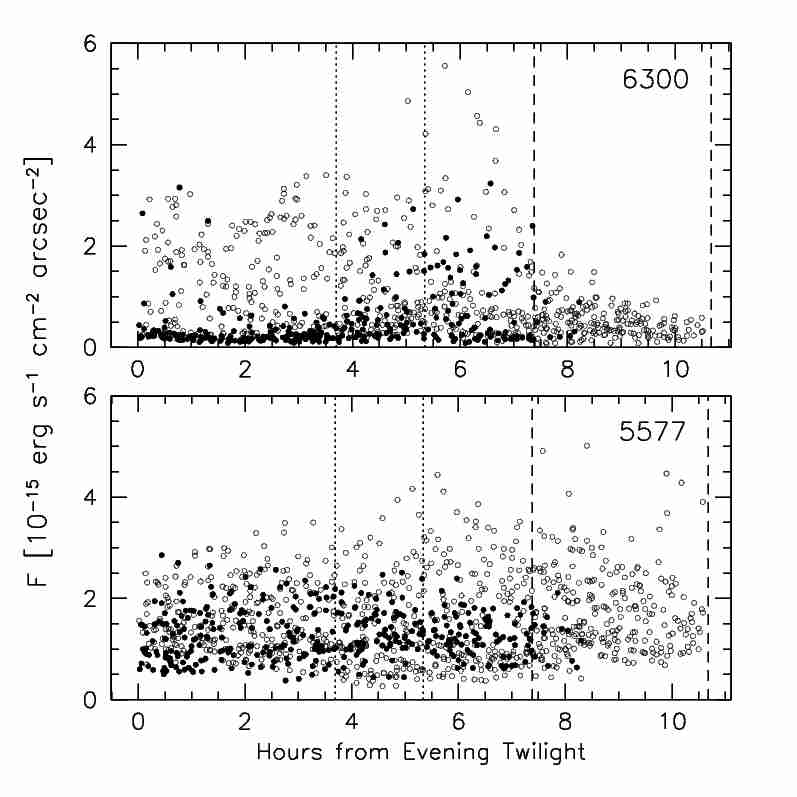}
\caption{\label{fig:decay}Line fluxes as a function of time elapsed from
evening twilight for [OI]6300 (upper panel) and [OI]5577
(lower panel). The vertical dashed lines mark the shortest and longest
nights (7.4 and 10.7 hours respectively), while dotted lines indicate
the midnight in the two cases. Filled symbols indicate data obtained
in nights whose length is between 7.4 and 8.4 hours.}
\end{figure}

It is well known that, during the first hours of the night, the
[OI]6300 line shows a steady decline in brightness which, at
tropical sites, is interrupted by abrupt emission bursts (see, for
instance, Roach \& Gordon \cite{roach}).  A plot of the line fluxes as
a function of time elapsed after the evening twilight (see
Fig.~\ref{fig:decay}) clearly shows that these events occur during the
whole night, and not only before midnight, as sometimes stated in the
literature (see for instance Benn \& Ellison \cite{benn}). Actually,
one interesting fact that emerges from this analysis is that, when the
nights are short (filled symbols), i.e. during austral summer, most of
the line enhancements tend to take place in the second half of the
night. On the other hand, when the nights are long (i.e. lasting more
than 8.5 hours), flux changes tend to become smaller and smaller
during the last two hours of the night.  No such dichotomy is observed
for the [OI]5577 line (see Fig.~\ref{fig:decay}, lower panel).

\begin{table}
\caption{\label{tab:lines} Linear least squares fit parameters for the
line fluxes vs.solar activity relation. Values in parenthesis indicate
RMS uncertainties. Input data have been corrected to zenith.}
\tabcolsep 2.2mm
\begin{tabular}{cccccc}
\hline
Line  & $\log F_0$ & $\gamma$ & $\sigma$ & $r$ & $N$ \\ 
\hline
OI~5577 & $-$15.36 (0.03) & 0.27 (0.02) & 0.22 & 0.49 & 876\\
OI~6300 & $-$15.69 (0.07) & 0.25 (0.05) & 0.41 & 0.23 & 635 \\
Na~I~D   & $-$15.72 (0.03) & 0.04 (0.02) & 0.25 & 0.07 & 876\\
NI~5200 & $-$17.19 (0.05) & 0.37 (0.03) & 0.32 & 0.45 & 727\\
\hline
\end{tabular}
\end{table}

\begin{figure}
\centering
\includegraphics[width=9cm]{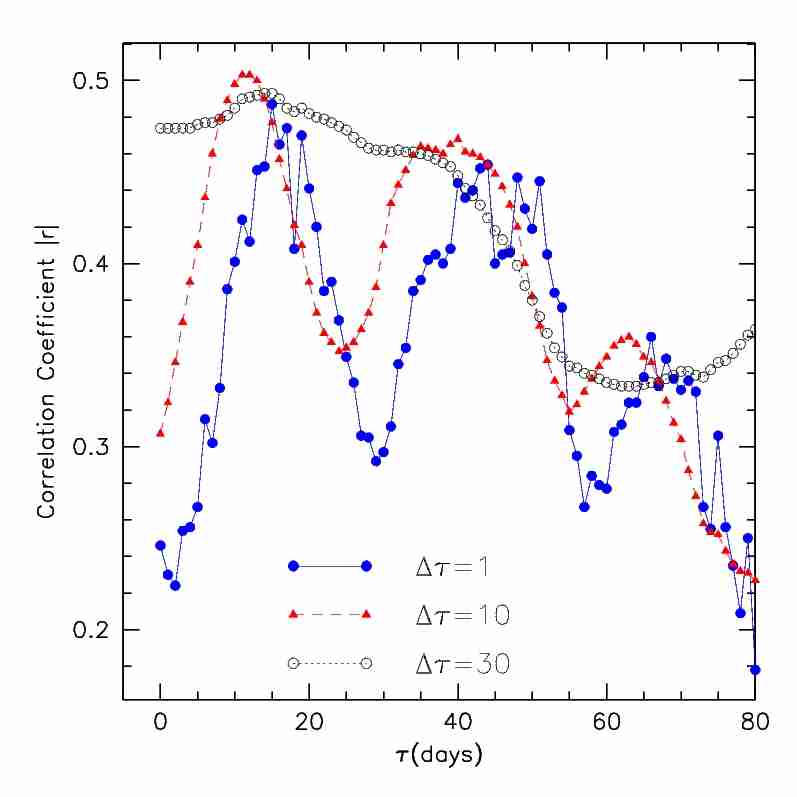}
\caption{\label{fig:scoi5577}Linear correlation coefficient (absolute value)
as a function of time delay $\tau$ for the [OI]5577 line,
computed for four different values of $\Delta \tau$ (1, 10, 30 days).}
\end{figure}

Finally, as for the broad band data (see Sec.~\ref{sec:timescale}), I
have calculated the correlation coefficient between line fluxes and
SFD for different values of the time delay $\tau$ and time window
$\Delta \tau$. An instructive example is presented in
Fig.~\ref{fig:scoi5577}, showing the case of the [OI]5577 line,
which displays the strongest dependency on solar activity. The
correlation function shows a peak at about $\tau$=15 days for $\Delta
\tau$=1 day. The maximum correlation increases for a $\Delta \tau$=10
days, peaking at $\tau\simeq$12.5 days. As for the photometric data
(see Fig.~\ref{fig:sunanal}), spurious correlation peaks due to the 27
days solar rotation are present. A similar analysis for the
[OI]6300 line shows two similar peaks at $\tau\simeq$13 days and
27 days later.

\subsubsection{\label{sec:naid}The Na~I D lines}

The Na~I~D doublet is known to originate in a layer placed at about 92
km and to undergo strong seasonal variations around an average value
of 50 R (Roach \& Gordon \cite{roach}). For the Paranal site, the
doublet varies by a factor larger than 15, reaching a maximum value of
160 R in April. The maximum contribution of the Na~I~D doublet to the
global night sky brightness reaches $\sim$0.1 mag arcsec$^{-2}$ both
in $V$ and $R$ passbands (see Table~\ref{tab:oi}).

\begin{figure}
\centering
\includegraphics[width=9cm]{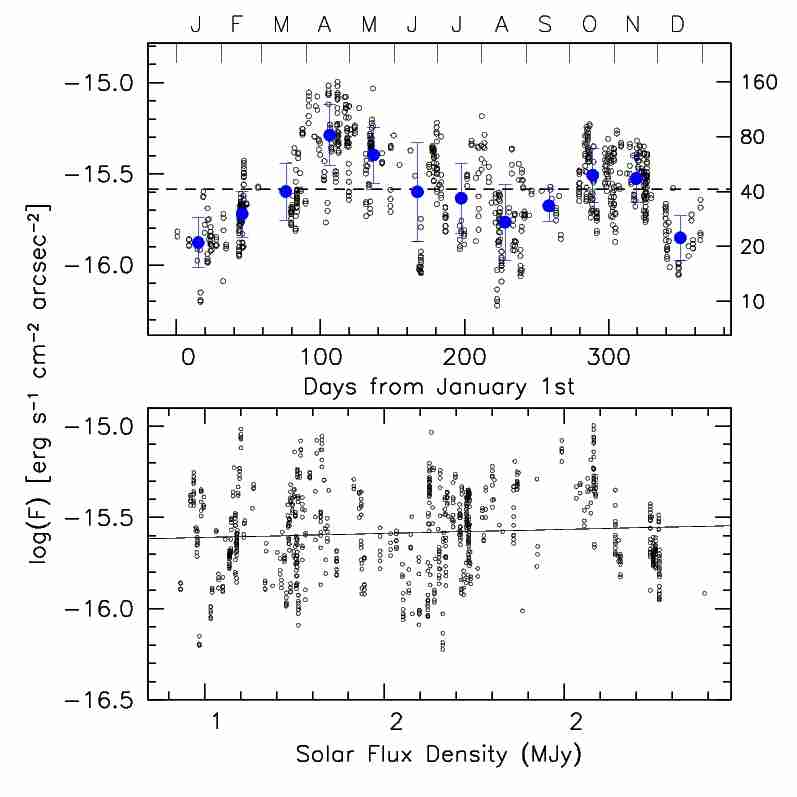}
\caption{\label{fig:naid}Same as Fig.~\ref{fig:oi5577} for the 
Na~I~D doublet. In this case, no solar flux correction has been applied.}
\end{figure}

While the Na~I~D doublet intensity does not show any significant
correlation with solar activity (the correlation coefficient is only
0.07; see Fig.~\ref{fig:naid} and Table~\ref{tab:lines}), it does show
a clear SAO, with a peak in April and a secondary peak in
October/November (see Fig.~\ref{fig:naid}, upper panel), as it is seen
in all atomic lines analyzed in this work. The SAO displayed by
mesospheric sodium column density is a well studied phenomenon
(Kirchoff \cite{kirkhoff}), also because of the importance of the
sodium layer for the laser guide star adaptive optics systems (see for
instance Ageorges \& Hubin \cite{ageorges}). The fact that different
species like Na~I and [OI] all show a SAO is interpreted in the light
of the role of O atoms in the source photochemical reactions (Slanger
\cite{slanger}).

\begin{figure}
\centering
\includegraphics[width=9cm]{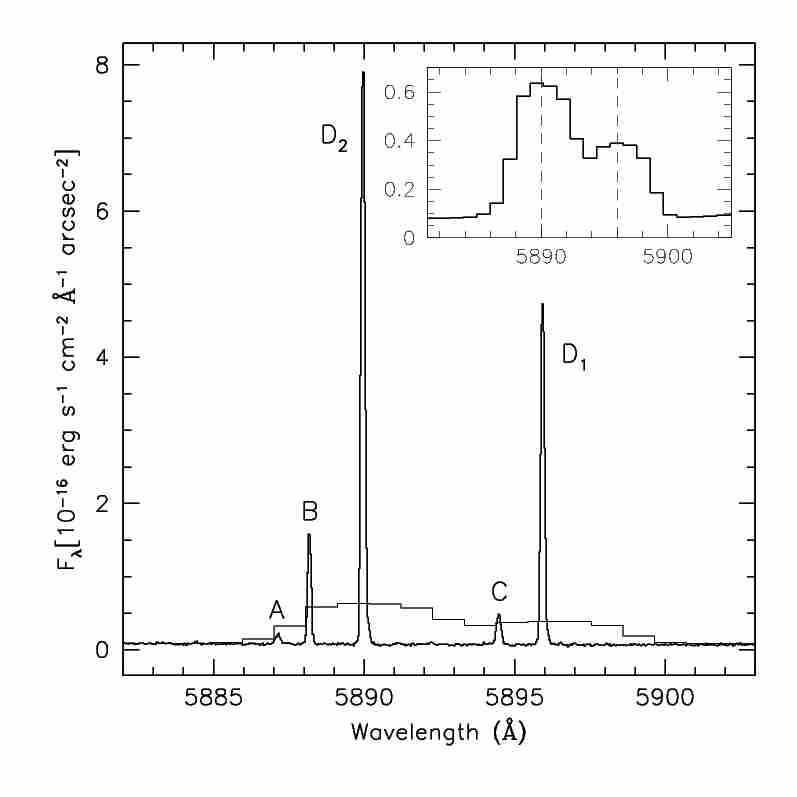}
\caption{\label{fig:uves}Night sky spectrum taken with UVES 
(Hanuschik \cite{hanuschik}) in the region of the Na~I~D lines
(resolution 0.15\AA\/ FWHM. The labels A, B and C indicate the OH
Meinel 8-2 band lines $Q_{22}(0.5)$ 5887.1\AA, $Q_{11}(1.5)$
5888.2\AA\/ and $Q_{11}(2.5)$ 5894.5\AA\/ (Abrams et
al. \cite{abrams}).  The insert shows the corresponding spectrum
obtained with FORS1 and the 600R grism (resolution 4.5\AA\/ FWHM).
The same spectrum is shown also in the main panel (thin line).}
\end{figure}

Since the resolution provided by the grism 600R is sufficient to
resolve the D$_1$ and D$_2$ components (see Fig.~\ref{fig:uves}),
following the work done by Slanger et al. (\cite{slanger}), I have
estimated the intensity ratio D$_2$/D$_1$ during dark time, on a total
of 147 spectra. The region of interest is contaminated by the presence
of at two features belonging to the OH(8-2) band, which I will
indicate as B and C (see Fig.~\ref{fig:uves} for their
identification). Using a high resolution UVES spectrum of the night
sky (Hanuschik \cite{hanuschik}), I have estimated the intensity ratio
between these two features and another OH(8-2) line at 5932.9\AA\/
($P_{11}(3.5)$, Abrams et al. \cite{abrams}), which I will indicate as
E. Since this feature is well measurable in the 600R spectra, assuming
that these intensity ratios are constant (B/E=0.95, C/B=0.35), I have
estimated the intensity of B and C from the measured intensity of
component E. Then, after subtracting to the original data two Gaussian
profiles centered at the laboratory wavelengths of B and C, I have
fitted the residual with a double Gaussian profile, fixing
the distance between the two components (5.98\AA) and their FWHM
(4.5\AA). Finally, I have derived the D$_2$/D$_1$ simply 
computing the ratio between the central intensities of the two
fitted Gaussians. The pseudo-continuum has been removed using a
first order polynomial.

\begin{figure}
\centering
\includegraphics[width=9cm]{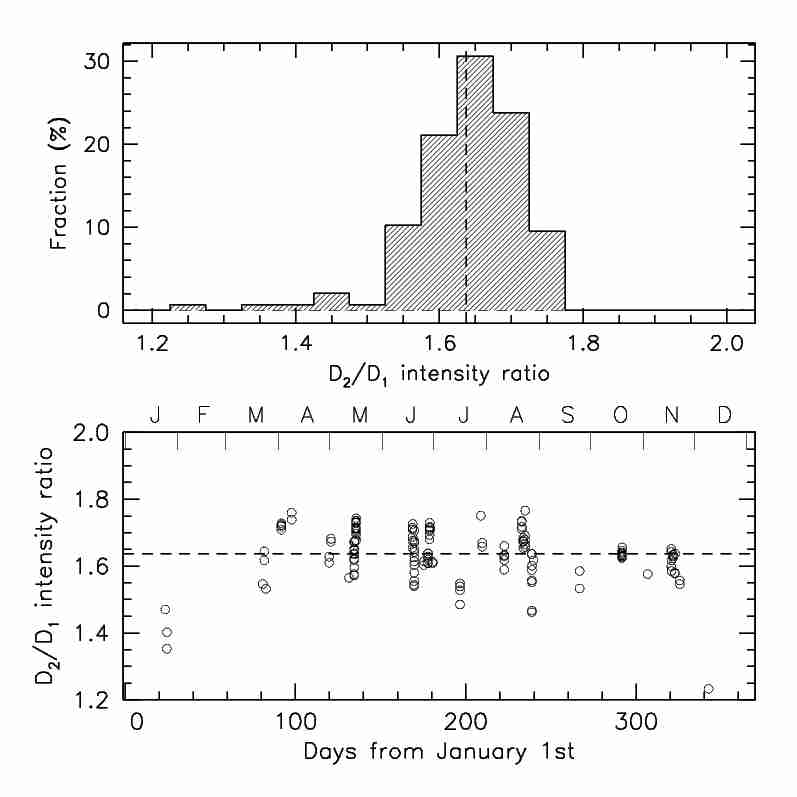}
\caption{\label{fig:naidratio}Upper panel: distribution of the Na~I D$_2$/D$_1$
intensity ratio during dark time. Lower panel: Na~I D$_2$/D$_1$
intensity ratio as a function of time from the beginning of the
year. The horizontal dashed line marks the average value.}
\end{figure}

The results, shown in Fig.~\ref{fig:naidratio} (upper panel), are
perfectly in line with the findings published by Slanger et
al. (\cite{slanger}): the intensity ratio, which should be 2.0 if the
two transitions are produced according to their spin-orbit statistical
weights, varies from 1.2 to 1.8, with most of the data lying between
1.5 and 1.7. The average value for the FORS1 sample is 1.64, with an
RMS deviation of 0.08.  Using a larger sample including more than 300
high resolution spectra, Slanger et al.  (\cite{slanger}) have found
evidences for a SAO for the D$_2$/D$_1$ intensity ratio. This is not
detected in the low-resolution data set presented here, but this is
probably due to the fact that the present sample includes less than
half as many data points.

\subsubsection{\label{sec:ni}The N~I 5200 line}

\begin{figure}
\centering
\includegraphics[width=9cm]{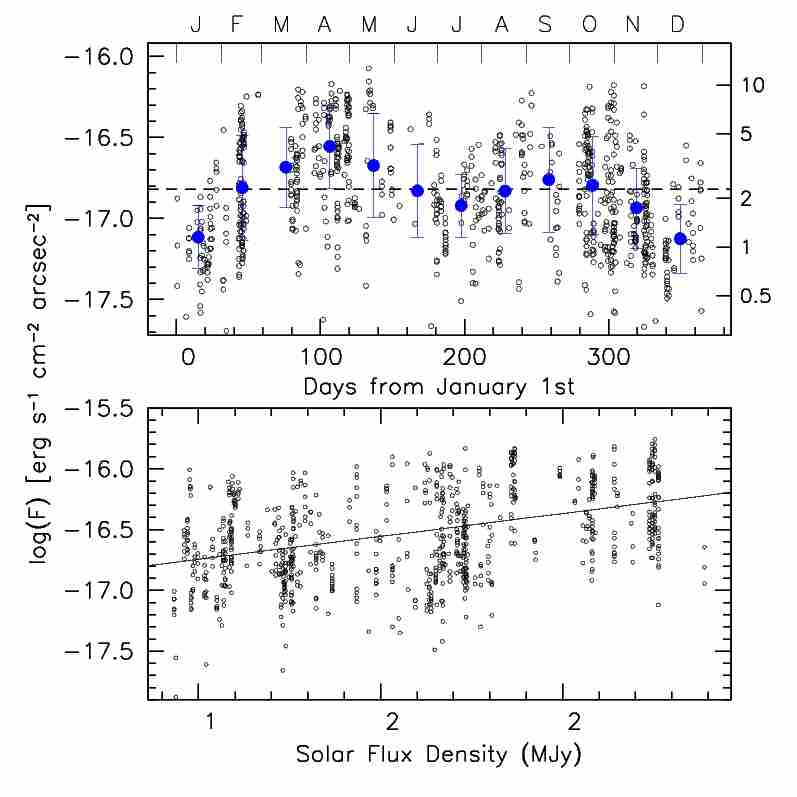}
\caption{\label{fig:ni5200}Same as Fig.~\ref{fig:oi5577} for the 
N~I~5200\AA\/ line.}
\end{figure}

The N~I feature at $\sim$5200\AA\/ is actually a blend of several
transitions.  It is supposed to originate at about 260 km altitude and
has a typical intensity of 1 R (Roach \& Gordon \cite{roach}). As I
have shown in Paper I, this line shows abrupt changes, possibly
following the behavior of the [OI]6300 line. The data discussed in
this paper show that the flux of this line ranges from practically
zero (the line is not detected) to about 30 R; moreover, the flux
distribution is rather similar to that of [OI]6300 (see
Fig.~\ref{fig:linehist}), strengthening the impression that these two
lines are related. This line displays also a
strong dependency on solar activity, with a correlation factor
similar to that of the [OI]5577 line (see Table~\ref{tab:lines}).
Finally, as for all other atomic lines discussed here, it shows a SAO
(see Fig.~\ref{fig:ni5200}).

The maximum contribution of this line to the global flux in the
$V$ band during dark time is about 2\%.

\subsection{\label{sec:bands}Main molecular emission bands}

\subsubsection{\label{sec:oh}The OH bands}

\begin{figure}
\centering
\includegraphics[width=9cm]{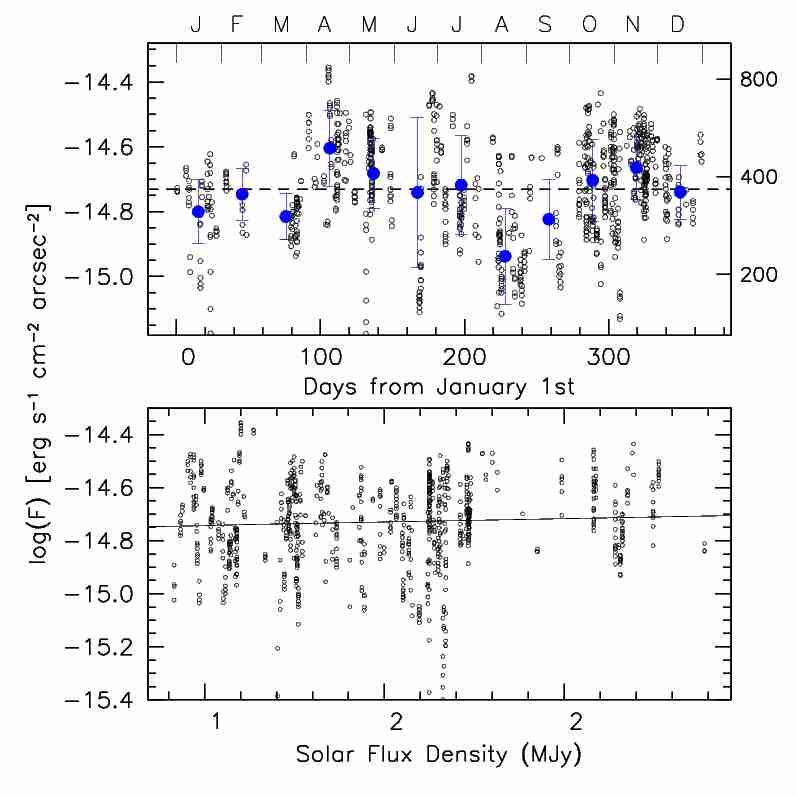}
\caption{\label{fig:oh83}Same as Fig.~\ref{fig:oi5577} for the 
OH(8-3) molecular band. No solar flux correction has been applied. }
\end{figure}

The OH bands analyzed here (see Table~\ref{tab:flux}) show a very
tight mutual correlation, in the sense that they appear to vary in
unison. Moreover, they do not show any correlation with solar activity
($r\leq$0.1 for all bands), while they show a SAO, even though not as
pronounced as in the case of the other features discussed so far (see
Fig.~\ref{fig:oh83} for an example). The flux distribution appears to
be much more symmetric around the average value than in the case of
atomic lines.  All bands shows the same range of variation, which is
close to a factor 2 around the mean value. Given the intensity of
these features (especially OH(8-3) and OH(6-2)), their variability is
certainly the dominating source of sky brightness fluctuations in the
I passband.

\subsubsection{\label{sec:02}The O$_2$(0-1) band}

\begin{figure}
\centering
\includegraphics[width=9cm]{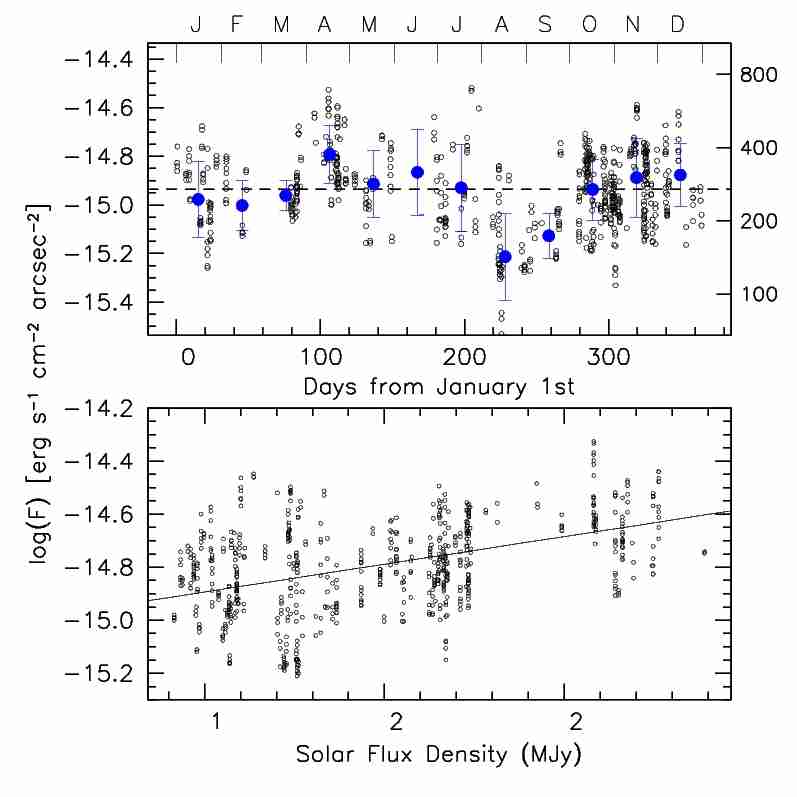}
\caption{\label{fig:o2}Same as Fig.~\ref{fig:oi5577} for the 
O$_2$(0-1) molecular band.}
\end{figure}

This band shows a clear correlation with solar activity ($r$=0.41) and
the same SAO observed for all other features (see Fig.~\ref{fig:o2}).
Its integrated flux varies by about a factor 15, between 80 R and 1.1
kR.

\subsection{\label{sec:blue} The blue bands and pseudo-continuum}

None of the blue bands, which are supposed to trace the behavior of
the pseudo-continuum generated by the O$_2$ and NO$_2$ molecular bands
(Roach \& Gordon \cite{roach}), shows a clear dependency on
solar activity ($r\leq$0.3). However, it must be noticed that, while
the emission features are purely generated within the atmosphere, the
continua are significantly influenced by the extra-terrestrial
background (Roach \& Gordon \cite{roach}), which is difficult to
remove. In fact, for the sake of simplicity, no differential zodiacal
light contribution has been applied to the continuum
measurements. This is certainly affecting the blue bands and the
continuum regions, especially those close to the zodiacal
light spectrum peak. This and the improper removal of airmass effects
is most likely the cause for the lower correlations shown by these
features and the solar activity, with respect to what is detected, for
example, for the $U$ and $B$ passbands.

\begin{figure}
\centering
\includegraphics[width=9cm]{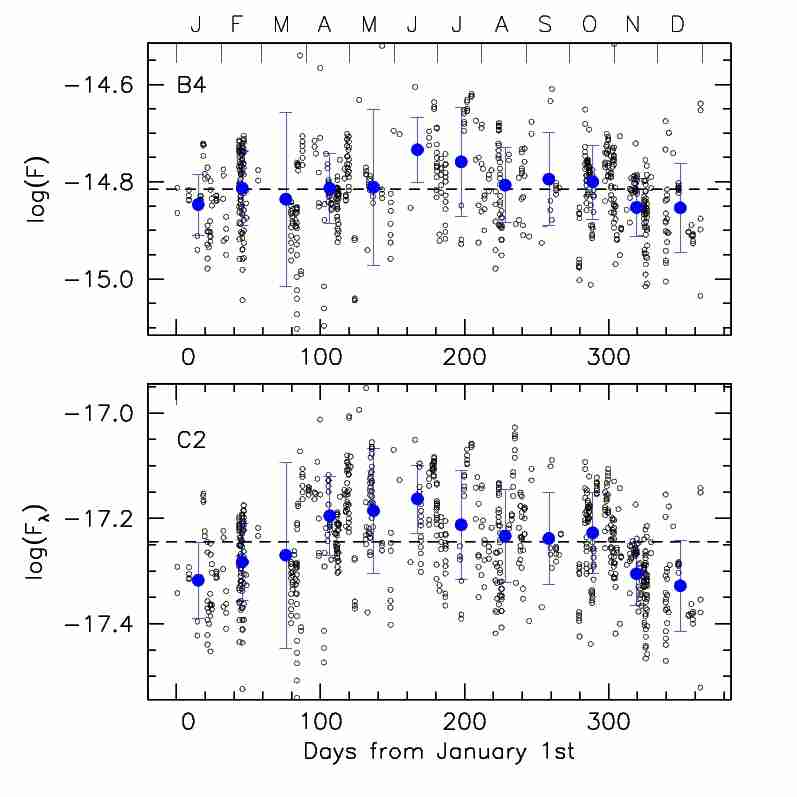}
\caption{\label{fig:b4}Seasonal variation of blue band B4 (upper panel) and
continuum range C2 (lower panel).}
\end{figure}

The integrated flux varies of about a factor 2 around the average
value, with a fairly symmetric distribution.  As already indicated by
the $B$ band data (see Sec.~\ref{sec:season}), there is no significant
trace of a SAO for none of the blue bands (see Fig.~\ref{fig:b4},
lower panel).  Interestingly, while also the continuum range C1
(5040-5120\AA, see Table~\ref{tab:flux}) does not display any evidence
for seasonal fluctuations, the redder ranges C2 to C6 show a possible
broad annual oscillation, with a peak in June (Fig.~\ref{fig:b4},
lower panel).

\begin{figure}
\centering
\includegraphics[width=9cm]{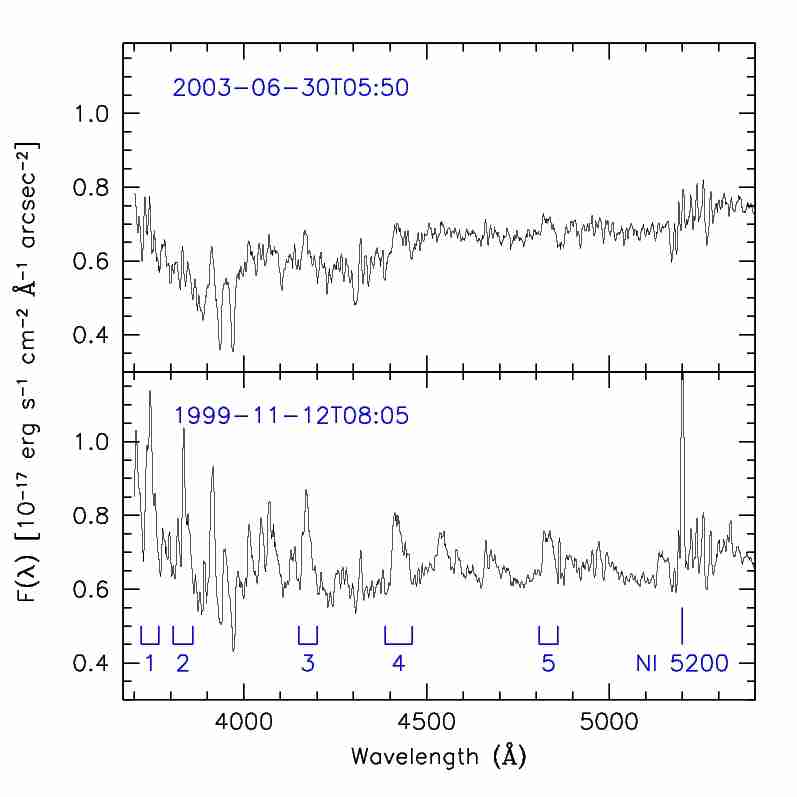}
\caption{\label{fig:blufeat}The spectral region 3750-5400\AA\/ on
two different epochs. Both spectra were obtained in dark time with the
600B grism. The two narrow absorption visible between 3900 and
4000\AA\/ are the CaII H\&K solar features.}
\end{figure}

While the region between 3100 and 3900\AA\/ (mostly not covered by the
data set discussed in this paper), shows a number of emission features
attributable to O$_2$ Herzberg and Chamberlain bands, the spectral
interval 3900-4900 is almost a pure pseudo-continuum (see for instance
Broadfoot \& Kendall \cite{broadfoot}, and Fig.~\ref{fig:bluband}
here). Nevertheless, clear variations are seen in the FORS1 database,
as it is illustrated in Fig.~\ref{fig:blufeat}, where I have compared
to spectra obtained with the same instrumental setup and similar
signal-to-noise ratio on two different nights. Whilst the two spectra
show the same overall emission features, these are much more
pronounced in the data obtained in November 1999 .

For an explorative analysis, I have measured the fluxes of some of the
most prominent features, which I have marked in Fig.~\ref{fig:blufeat}
(lower panel) and I will indicate as bf1 (3720-3767\AA), bf2
(3807-3860\AA), bf3 (4150-4200\AA), bf4 (4387-4460\AA) and bf5
(4809-4859\AA). While bf1, bf2 and bf3 can be tentatively identified
as N$_2$ and N$^+_2$ molecular bands (see Chamberlain
\cite{chamberlain} and also the next section here), the identification of 
bf4 and bf4 is more difficult.  For simplicity, I have used a common
value for the blue pseudo-continuum (CB), which was estimated in the
spectral range 4730-4760\AA, that does not show any trace of emission
features (see for instance Hanuschik \cite{hanuschik}). The results
are summarized in Table~\ref{tab:blufeat}, that reports the maximum
and median values derived from the 114 suitable spectra used in the
analysis, together with the percentage of non-detections.

The largest range of variation is shown by bf1, which can reach a
maximum flux of about 1.4$\times$10$^{-16}$ erg s$^{-1}$ cm$^{-2}$
arcsec$^{-2}$, while bf2 to bf5 reach peak fluxes between 6 and
8$\times$10$^{-17}$ erg s$^{-1}$ cm$^{-2}$ arcsec$^{-2}$. Despite
their similar peak values, these features display different
behaviors. For instance, while bf5 is practically always detected, bf2
is absent in more than 50\% of the cases. Since the median value of
the pseudo-continuum is 7.2$\times$10$^{-18}$ erg s$^{-1}$ cm$^{-2}$
arcsec$^{-1}$, each of these features contributes less than 2\% to the
integrated flux between 3700 and 5000\AA. Therefore, their
fluctuations can account only partially for the variations observed in
the B passband, that must be related to the changes in the
pseudo-continuum.

\begin{table}
\tabcolsep 2.5mm
\caption{\label{tab:blufeat} Dark time maximum and median fluxes for the 
blue emission features indicated in Fig.~\ref{fig:blufeat}. The last
column gives the frequency of non-detection (flux below 10$^{-17}$ erg
s$^{-1}$ cm$^{-2}$ arcsec$^{-1}$).}
\centerline{
\begin{tabular}{ccccc}
\hline
bf   &  wavelength range        & max  &  med  &  n.d. \\
\#   & (\AA)         &
\multicolumn{2}{c}{(10$^{-17}$ erg s$^{-1}$ cm$^{-2}$ arcsec$^{-2}$}) & \\
\hline
1    & 3720-3767     & 14.4 & 5.4   &  26\% \\
2    & 3807-3860     &  8.4 & 0.0   &  54\% \\
3    & 4150-4200     &  6.4 & 1.3   &  38\% \\
4    & 4387-4460     &  7.2 & 2.8   &  17\% \\
5    & 4809-4859     &  8.0 & 3.9   &   1\% \\
\hline
\end{tabular}
}
\end{table}

\subsubsection{The strange event of November 8, 2004}

\begin{figure}
\centering
\includegraphics[width=9cm]{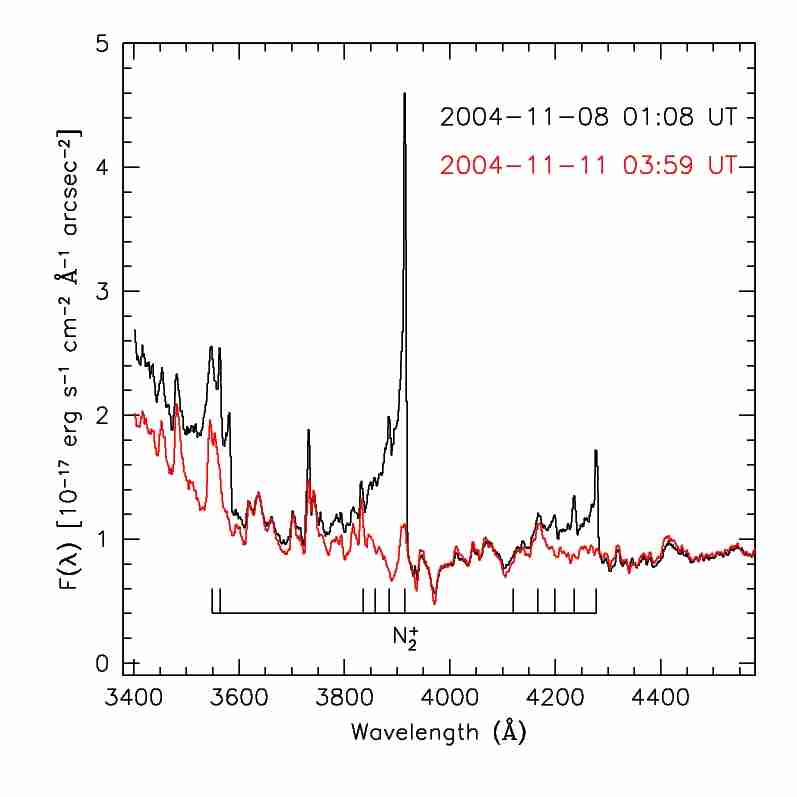}
\caption{\label{fig:thunder}The weird night sky spectrum of November 8, 2004.
The N$_2^+$ first negative band identifications are from Chamberlain
(\cite{chamberlain}; Table~5.4). For presentation the Nov 11 spectrum
has been scaled by a factor 1.1, in order to match the
pseudo-continuum on Nov 8.}
\end{figure}

Normally, there are no conspicuous isolated emission features in the
airglow bluewards of 5200\AA, which is dominated by the
pseudo-continuum (Roach \& Gordon \cite{roach}. See also
Fig.~\ref{fig:bluband} here). However, N. Castro and M. Garcia, while
analyzing a set of low resolution spectroscopic data obtained with
FORS2 on November 8, 2004, have noticed the presence of unexplained
emission features between 3600 and 4400\AA. A more detailed analysis
has shown that these emissions where present on all the FORS2 data of
that night, i.e. a set of MXU exposures 2700 seconds each, obtained
between 00:15 and 07:08 UT\footnote{On November 8, 2004 the evening
twilight ended on 00:23 UT and the morning twilight started on 08:27
UT.} using the 600B grism. As expected, no trace of these features was
visible in a similar data set obtained three days later. An
example is shown in Fig.~\ref{fig:thunder} where, for comparison, an
analogous spectrum obtained on November 11 is also plotted. Clearly,
the two spectra differ mainly for the presence of two prominent
emission bands, peaking at 3194\AA\/ and 4278\AA, which are identified
as N$_2^+$ first negative bands 1N(0-0) and 1N(0-1) (Chamberlain
\cite{chamberlain}; Table~5.4). These features, which are normally
very weak or even absent in the nightglow (Broadfoot \& Kendall
\cite{broadfoot}) are on the contrary typical of the aurora spectrum
(Chamberlain
\cite{chamberlain}). Besides being an extremely strange phenomenon at
the latitudes of Paranal, an aurora would certainly be accompanied by
other spectral markers, like for instance a large increase in the
emission of the [OI]5577 line, which can reach in fact an
intensity of 100 kR during a IBC III aurora (Chamberlain
\cite{chamberlain}). The flux carried by this line in the same spectrum
presented in Fig.~\ref{fig:thunder} is $\sim$183 R, that is slightly
below the average level measured for Paranal (230R, see
Sec.~\ref{sec:oi5577}). This definitely rules out an exceptional
auroral event as the responsible for the unusual spectrum observed on
November 8, 2004.

A plausible explanation, proposed by Castro \& Garcia (private
communication), is the contamination by the reflection from clouds of
a number of lightning strokes. In effects, in the spectral range
covered by the FORS2 data (3600-6100\AA), the most prominent features
in a lightning spectrum are the N$_2^+$ first negative bands 1N(0-0)
and 1N(0-1) (see for instance Wallace \cite{wallace}). Additionally,
on the night of Nov 8 2004 thick and thin cirrus were reported in the
ESO-Paranal night logs, substantiating the hypothesis of scattered
light from a rather far thunderstorm.

This kind of events must be indeed very rare, since no other example
could be found in the FORS1 spectral data base presented in this
paper.

\section{\label{sec:corr}Correlations between spectral features}

For a first exploratory analysis I have computed the linear
correlation coefficients in the $\log F$-$\log F$ plane between all
measured features.. The results are presented in
Table~\ref{tab:fluxcorr} and they basically confirm the correlations
found by Barbier (\cite{barbier}), even though new interesting facts
do appear.

\begin{figure}
\centering
\includegraphics[width=9cm]{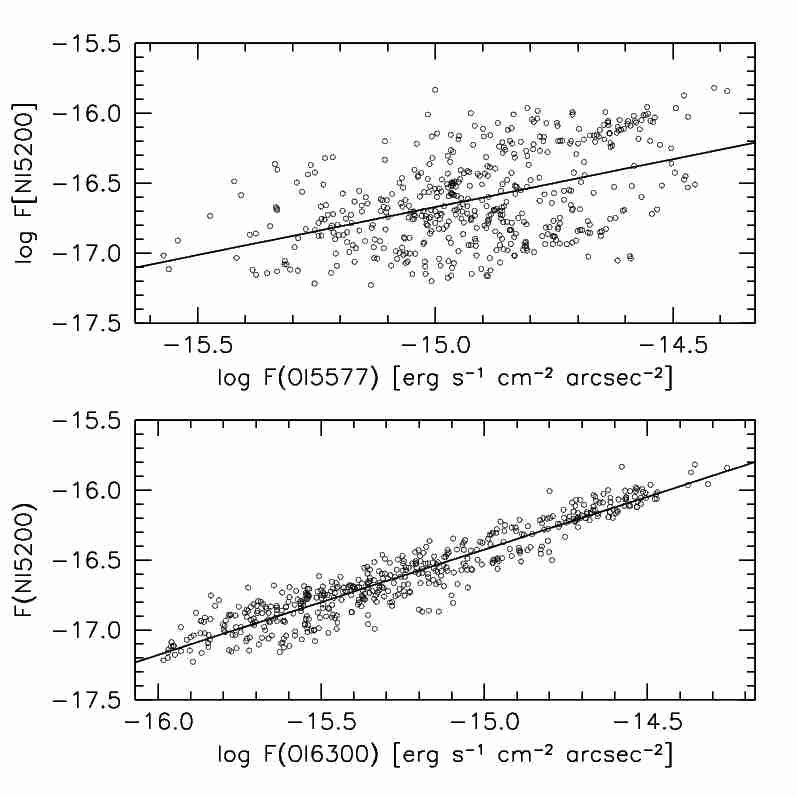}
\caption{\label{fig:oinicorr}Correlation between N~I 5200 and
[OI]5577 (upper panel) and [OI]6300 (lower panel). Only
airmass correction has been applied.}
\end{figure}

The pioneering optical, eight-color photometric studies by Barbier
(\cite{barbier}) have shown the existence of the so-called {\it
covariance groups}: the {\bf green-line group} ([OI]5577, O$_2$
Herzberg bands, the blue bands, the green continuum and the O$_2$(0-1)
band), the {\bf sodium group} (Na~I~D doublet and the OH bands) and the
{\bf red-line group}, which includes only the [OI]6300,6364
doublet). So far, the latter appeared to be completely independent
from any other component of the airglow (see Chamberlain
\cite{chamberlain}). Nevertheless, as it is shown in 
Fig.~\ref{fig:oinicorr}, [OI]6300 shows a very tight correlation
with the N~I feature at 5200\AA. The linear correlation factor in the
log-log plane is $r$=0.95 and this appears to be one of the strongest
correlation between airglow features found in the data set presented
here, surpassed only by that shown by the OH bands (see
Table~\ref{tab:fluxcorr}). To my knowledge, this is the first time
this finding is reported; most likely, it escaped the attention of
previous investigations simply because the N~I feature is rather weak
($\leq$30 R) and hence practically impossible to measure with
intermediate passband filters.

Even though a correlation between N~I 5200 and [OI]5577 is
found (see Fig.~\ref{fig:oinicorr}, upper panel), this is less marked
($r$=0.56) and the spread around the best fit relation is much larger
($\sigma$=0.29 vs. $\sigma$=0.11). Finally, the correlation between
[OI]5577 and [OI]6300 is indeed weak ($r$=0.29,
$\sigma$=0.39); nevertheless, the data presented here seem to indicate
that, on average, the maximum value attained by the red line is
related to the flux of the green-line through the simple relation
F([OI]6300)$\leq$2$\times$F([OI]5577).

\begin{table*}
\caption{\label{tab:fluxcorr}Linear correlation factors in 
the log F-log F plane. Only dark time data have been used. The last
two columns and rows report the correlation with daily (SFD$_d$) and
monthly (SFD$_m$) solar flux density averages in the semi-log plane.}
\tabcolsep 0.8mm
\begin{tabular}{l|cccccccccccccccccccc|cc}
\hline
&C1 &5200 &C2   & Na~ID & C3  &5577 &C4  &6300 &C5 &(6-1) &(7-2) &(8-3) &C6 &(6-2)
& O$_2$ &B1 &B2 &B3 &B4 &Broad & SFD$_d$ &SFD$_m$ \\
\hline
C1     &     &0.15 &0.79 &0.14 &0.94 &0.31 &0.82 &0.02 &0.88 &0.23 &0.14 &0.15 &0.67 &0.14 &0.16 &0.87 &0.95 &0.99 &1.00 &0.60 &0.25 &0.23\\
5200   &0.15 &     &0.36 &0.43 &0.29 &0.56 &0.31 &0.95 &0.22 &0.34 &0.32 &0.32 &0.35 &0.29 &0.21 &0.31 &0.22 &0.16 &0.17 &0.60 &0.31 &0.45\\   	
C2     &0.79 &0.36 &     &0.58 &0.94 &0.41 &0.99 &0.28 &0.96 &0.40 &0.32 &0.37 &0.92 &0.46 &0.35 &0.75 &0.77 &0.76 &0.79 &0.90 &0.12 &0.14\\    
Na~ID   &0.14 &0.43 &0.58 &     &0.35 &0.41 &0.52 &0.32 &0.41 &0.70 &0.72 &0.75 &0.63 &0.61 &0.61 &0.19 &0.14 &0.15 &0.16 &0.72 &0.03 &0.07\\  
C3     &0.94 &0.29 &0.94 &0.35 &     &0.43 &0.97 &0.17 &0.98 &0.21 &0.12 &0.16 &0.86 &0.33 &0.32 &0.85 &0.89 &0.92 &0.94 &0.80 &0.23 &0.22\\ 
5577   &0.31 &0.56 &0.41 &0.41 &0.43 &     &0.38 &0.29 &0.31 &0.31 &0.29 &0.32 &0.26 &0.25 &0.85 &0.61 &0.55 &0.37 &0.36 &0.59 &0.30 &0.49\\  
C4     &0.82 &0.31 &0.99 &0.52 &0.97 &0.38 &     &0.26 &0.99 &0.36 &0.28 &0.32 &0.93 &0.44 &0.33 &  -  &  -  &0.81 &0.83 &0.90 &0.13 &0.13\\  
6300   &0.02 &0.95 &0.28 &0.32 &0.17 &0.29 &0.26 &     &0.19 &0.15 &0.13 &0.13 &0.28 &0.15 &0.01 &  -  &  -  &0.02 &0.02 &0.49 &0.09 &0.23\\  
C5     &0.88 &0.22 &0.96 &0.41 &0.98 &0.31 &0.99 &0.19 &     &0.27 &0.19 &0.23 &0.91 &0.37 &0.26 &  -  &  -  &0.87 &0.88 &0.83 &0.14 &0.10\\  
(6-1)  &0.23 &0.34 &0.40 &0.70 &0.21 &0.31 &0.36 &0.15 &0.27 &     &0.98 &0.97 &0.67 &0.97 &0.53 &  -  &  -  &0.25 &0.25 &0.61 &0.02 &0.07\\  
(7-2)  &0.14 &0.32 &0.32 &0.72 &0.12 &0.29 &0.28 &0.13 &0.19 &0.98 &     &1.00 &0.63 &0.99 &0.57 &  -  &  -  &0.17 &0.17 &0.55 &0.06 &0.04\\  
(8-3)  &0.15 &0.32 &0.37 &0.75 &0.16 &0.32 &0.32 &0.13 &0.23 &0.97 &1.00 &     &0.65 &0.98 &0.60 &  -  &  -  &0.18 &0.18 &0.59 &0.04 &0.06\\  
C6     &0.67 &0.35 &0.92 &0.63 &0.86 &0.26 &0.93 &0.28 &0.91 &0.67 &0.63 &0.65 &     &0.64 &0.34 &  -  &  -  &0.68 &0.68 &0.90 &0.02 &0.05\\  
(6-2)  &0.14 &0.29 &0.46 &0.61 &0.33 &0.25 &0.44 &0.15 &0.37 &0.97 &0.99 &0.98 &0.64 &     &0.47 &  -  &  -  &0.16 &0.16 &0.63 &0.06 &0.10\\  
O$_2$  &0.16 &0.21 &0.35 &0.61 &0.32 &0.85 &0.33 &0.01 &0.26 &0.53 &0.57 &0.60 &0.34 &0.47 &     &  -  &  -  &0.24 &0.22 &0.52 &0.22 &0.41\\  
B1     &0.87 &0.31 &0.75 &0.19 &0.85 &0.61 &  -  &  -  &  -  &  -  &  -  &  -  &  -  &  -  &  -  &     &0.96 &0.91 &0.89 &  -  &0.28 &0.25\\  
B2     &0.95 &0.22 &0.77 &0.14 &0.89 &0.55 &  -  &  -  &  -  &  -  &  -  &  -  &  -  &  -  &  -  &0.96 &     &0.99 &0.97 &  -  &0.28 &0.24\\  
B3     &0.99 &0.16 &0.76 &0.15 &0.92 &0.37 &0.81 &0.02 &0.87 &0.25 &0.17 &0.18 &0.68 &0.16 &0.24 &0.91 &0.99 &     &0.99 &0.61 &0.24 &0.25\\			            
B4     &1.00 &0.17 &0.79 &0.16 &0.94 &0.36 &0.83 &0.02 &0.88 &0.25 &0.17 &0.18 &0.68 &0.16 &0.22 &0.89 &0.97 &0.99 &     &0.62 &0.27 &0.26\\  
Broad  &0.60 &0.60 &0.90 &0.72 &0.80 &0.59 &0.90 &0.49 &0.83 &0.61 &0.55 &0.59 &0.90 &0.63 &0.52 &  -  &  -  &0.61 &0.62 &     &0.14 &0.24\\  
\hline
SFD$_d$&0.25 &0.31 &0.12 &0.03 &0.23 &0.30 &0.13 &0.09 &0.14 &0.02 &0.06 &0.04 &0.02 &0.06 &0.22 &0.28 &0.28 &0.24 &0.27 &0.14 &     &0.71\\
SFD$_m$&0.23 &0.45 &0.14 &0.07 &0.22 &0.49 &0.13 &0.23 &0.10 &0.07 &0.04 &0.06 &0.05 &0.10 &0.41 &0.25 &0.24 &0.25 &0.26 &0.24 &0.71 &    \\         
\hline
\end{tabular}
\end{table*}

The strongest correlation within the [OI]5577 covariance group is
that with the O$_2$ band ($r$=0.85), followed by the blue bands B1
($r$=0.61), N~I 5200\AA\/ ($r$=0.56), B2 ($r$=0.55), C3 ($r$=0.43) and
C2 ($r$=0.41). The correlation with the other continuum regions is
weaker ($r\leq$0.4). As for the Na~I~D group, besides the very tight
correlations existing between the OH bands ($r\geq$0.97), the
strongest correlation is observed between Na~I~D and OH(8-3)
($r$=0.75), followed by the other OH bands. Na~I~D correlates rather
well with the continuum ranges C2 ($r$=0.58) and C6 ($r$=0.63) and
with the O$_2$ band ($r$=0.61). As anticipated in the previous
section, the only meaningful correlation found in the third covariance
group is between [OI]6300 and the N~I 5200 line
($r$=0.95). This is actually one of the tightest correlations found
among all features. Finally, all continuum bands C1 to C6 are well
correlated with each other and with the blue bands B1 to B4, which
show as well a very strong mutual correlation.

Even though the analysis is certainly hampered by the smaller sample,
I have run a similar exploratory study for the blue emission features
described in Sec.~\ref{sec:blue}. The results are presented in
Table~\ref{tab:blucorr}. The first interesting fact is that the blue
emission features bf1 to bf4 are uncorrelated with the CB continuum
($|r|\leq$0.15), while some correlation is seen for bf5
($r\simeq$0.4). On the other hand, all blue features show a
significant mutual correlation, which ranges from a minimum
($r\simeq$0.4) for bf2 and bf5 to a maximum ($r\simeq$0.9) for bf3 and
bf4. In general, bf5 is the feature that shows the weakest correlation
with the remaining blue features. Additionally, it displays the
strongest correlation with the [OI]~5577 feature ($r\simeq$0.8) and
solar activity. Indeed, the correlation with [OI]~5577 is rather
marked for all bf's ($r\geq$0.5), suggesting that these features might
belong to the green-line group of Barbier
(\cite{barbier}). Nevertheless, bf1 and bf2 show a similarly marked
correlation to the N~I 5200 line ($r\geq$0.5), that belongs to the
red-line group. This suggests a partial correlation between the abrupt
micro-auroral events undergone by the [OI]6300,6364\AA\/ doublet and
the blue features activity.

\begin{table*}
\caption{\label{tab:blucorr}Same as Table~\ref{tab:fluxcorr} for the blue features.}
\tabcolsep 0.8mm
\centerline{
\begin{tabular}{l|cccccccccc|cc}
\hline
  &C1 &5200 & C3 &5577 & CB & bf1 & bf2 & bf3 & bf4 & bf5 & SFD$_d$
  &SFD$_m$ \\
\hline
C1   &     &0.21 &0.97 &0.40 &1.00 &0.05 &0.01 &0.15 &0.15 &0.40 &0.32 &0.30\\
5200 &0.21 &     &0.32 &0.59 &0.20 &0.56 &0.51 &0.21 &0.12 &0.13 &0.49 &0.46\\
C3   &0.97 &0.32 &     &0.48 &0.96 &0.15 &0.19 &0.23 &0.07 &0.46 &0.38 &0.36\\
5577 &0.40 &0.59 &0.48 &     &0.38 &0.65 &0.47 &0.54 &0.56 &0.80 &0.54 &0.45\\
CB   &1.00 &0.20 &0.96 &0.38 &     &0.05 &0.00 &0.15 &0.14 &0.38 &0.30 &0.28\\
bf1  &0.05 &0.56 &0.15 &0.65 &0.05 &     &0.82 &0.76 &0.54 &0.50 &0.39 &0.30\\
bf2  &0.01 &0.51 &0.19 &0.47 &0.00 &0.82 &     &0.80 &0.69 &0.44 &0.32 &0.08\\
bf3  &0.15 &0.21 &0.23 &0.54 &0.15 &0.76 &0.80 &     &0.91 &0.62 &0.21 &0.02\\
bf4  &0.15 &0.12 &0.07 &0.56 &0.14 &0.54 &0.69 &0.91 &     &0.69 &0.26 &0.18\\
bf5  &0.40 &0.13 &0.46 &0.80 &0.38 &0.50 &0.44 &0.62 &0.69 &     &0.50 &0.32\\
\hline
SFD$_d$ &0.32 &0.49 &0.38 &0.54 &0.30 &0.39 &0.32 &0.21 &0.26 &0.50 &    &0.79\\
SFD$_m$ &0.30 &0.46 &0.36 &0.45 &0.28 &0.30 &0.08 &0.02 &0.18 &0.32 &0.79 &    \\
\hline
\end{tabular}
}
\end{table*}

As far as the correlation with solar activity is concerned, it is
interesting to note that while the blue pseudo-continuum shows a very
similar correlation factor with $SFD_d$ and $SFD_m$, for the blue
features bf2 and bf3 this is significantly larger when the daily
averages are used.  On the contrary, bf5 shows a stronger correlation
with the monthly averages. Because of the limited sample, though,
these results have to be taken with some caution. In fact, for
instance, the [OI]5577 line shows a suspiciously high correlation
factor with $SFD_d$ (0.54), at variance with the value derived from
the whole data set (0.30; see Table~\ref{tab:fluxcorr}).

\section{\label{sec:short}Short timescale variations}

\begin{figure}
\centering
\includegraphics[width=9cm]{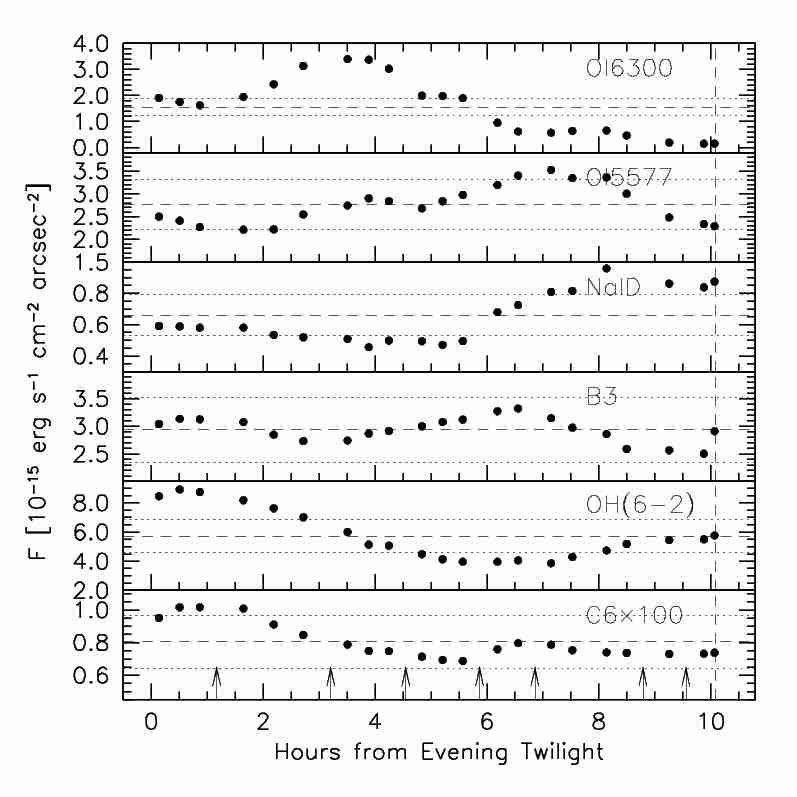}
\caption{\label{fig:seq}Sequence of spectral measurements for some of
the features studied in this work. The data were obtained with FORS1
on 2001-04-22 (G300V+GG435). Exposure times ranged from 10 to 20
minutes. Several sky targets at different airmasses were observed. The
vertical arrows indicate the times of target change. Airmasses spanned
from 1.2 to 1.9. The vertical dashed line marks the start of morning
astronomical twilight. The horizontal dashed lines mark the average
value, while the dotted lines indicate $\pm$20\% levels.}
\end{figure}

In general, the spectroscopic time coverage of the present sample is
sparse and this makes the study of short time scale variations
(minutes to hours) rather difficult. Interestingly, the sample
presented here includes some data sets for which the spectroscopic
observations went on for a significant fraction of the night. The most
complete case is illustrated in Fig.~\ref{fig:seq}, which presents the
behavior of some selected features during the night of 2001-04-22 as a
function of time elapsed from the end of evening twilight, with an
average interval of thirty minutes. This sequence definitely shows that
the variations are smooth, with typical timescales of the order of
several hours. As expected, the most marked change is seen in the
[OI]6300 feature that, during the first hours of the night, went
through one of the intensity enhancements typical for sites close to
the geomagnetic equator.

The smooth time evolution seen in the emission features matches the
behavior observed in broad band photometry when long time series are
available (see Paper I, Sec.~6).

\section{\label{sec:disc}Discussion}

The analysis presented in this paper gives a clear picture of the
complexity shown by the nightglow fluctuations, most of which remain
unexplained. If it is well established that several features show a
definite correlation with solar activity, to which they react with
timescales of the order of a couple of weeks, more thorough
investigations need to be performed in order to better understand the
link between space weather and the phenomena taking part in the upper
layers of Earth's atmosphere.

In all the studies of the nightglow in the astronomical context, the
radio flux at 10.2 cm has been used as the only proxy for the solar
activity. Nevertheless, other transient solar phenomena might have
some impact on the night sky brightness, like Flares, Coronal Mass
Ejections and Solar Proton Events (see Hanslmeier
\cite{hanselmeier} for a very recent review on solar phenomena).
During these events, large amounts of energetic charged particles are
released and, in the case they interact with the Earth's magnetic
field, they cause a series of geomagnetic effects, including aurorae.
Therefore, charged particles are potential responsibles for at least
some of the observed nightglow fluctuations, including the SAO.

\begin{figure}
\centering
\includegraphics[width=9cm]{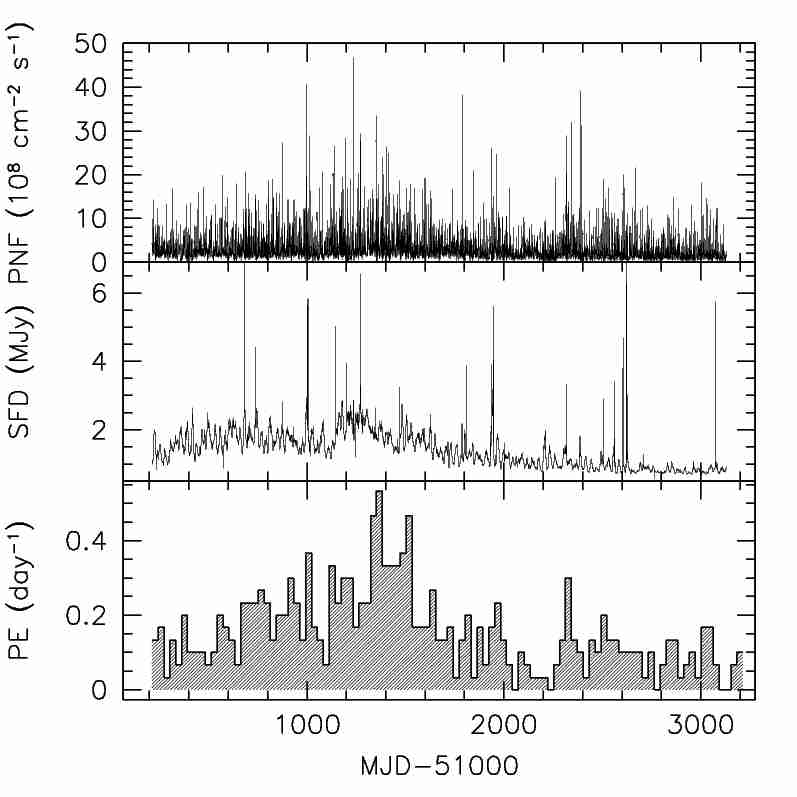}
\caption{\label{fig:sohospot}Upper panel: CELIAS/MTOF proton number flux.
Central panel: solar radio flux. Lower panel: CELIAS/MTOF proton
events (PF$\geq$7$\times$10$^8$ cm$^{-2}$ s$^{-1}$ ).}
\end{figure}

Even though this will require a dedicated analysis, I have run a
preliminary study using the proton flux measured by the CELIAS/MTOF
Proton Monitor (Ipavich et al. \cite{ipavich}) on board of
SOHO\footnote{Data can be downloaded from
http://umtof.umd.edu/pm/crn/}, orbiting on the Earth-Sun line at about
1.5$\times$10$^6$ km away from Earth. The proton number flux (PNF) has
been derived multiplying the proton density by the proton velocity
measured by CELIAS/MTOF and it is plotted in Fig.~\ref{fig:sohospot}
for the relevant time interval (upper panel).

The PNF average value during this time span is 2.6$\times$10$^8$
cm$^{-2}$ s$^{-1}$ (corresponding to a mass loss rate of
$\sim$2$\times$10$^{-14}$ M$_\odot$ yr$^{-1}$), while the yearly
averages range from 1.9 to 3.5$\times$10$^8$ cm$^{-2}$ s$^{-1}$. The
maximum average value was reached in July 2002, i.e. some time after
the secondary maximum seen in the radio flux. The proton velocity
ranges from 270 to 1000 km s$^{-1}$ (median value 440 km s$^{-1}$),
while the number density ranges from 0.1 to 73 cm$^{-3}$ (median value
4.7 cm$^{-3}$). Looking in more detail at the SOHO data, one notices
that a number of isolated and short duration peaks are present in the
PNF. In fact, in some cases, values as high as 4$\times$10$^9$
cm$^{-2}$ s$^{-1}$ are reached. Already a look at
Fig.~\ref{fig:sohospot} (upper panel) shows that the frequency of
these spikes is higher closer to the solar maximum. This can be seen
more quantitatively in the bottom panel of Fig.~\ref{fig:sohospot},
where I have plotted the rate of what I will indicate as proton events
(PE)\footnote{Note that these events differ from the Solar Proton
Events, which are bursts of relativistic protons, with energies larger
than 10 MeV.}. In this context, a PE has been defined as a group of
adjacent CELIAS/MTOF measurements with PNF$\geq$7$\times$10$^8$
cm$^{-2}$ s$^{-1}$.

\begin{figure}
\centering
\includegraphics[width=9cm]{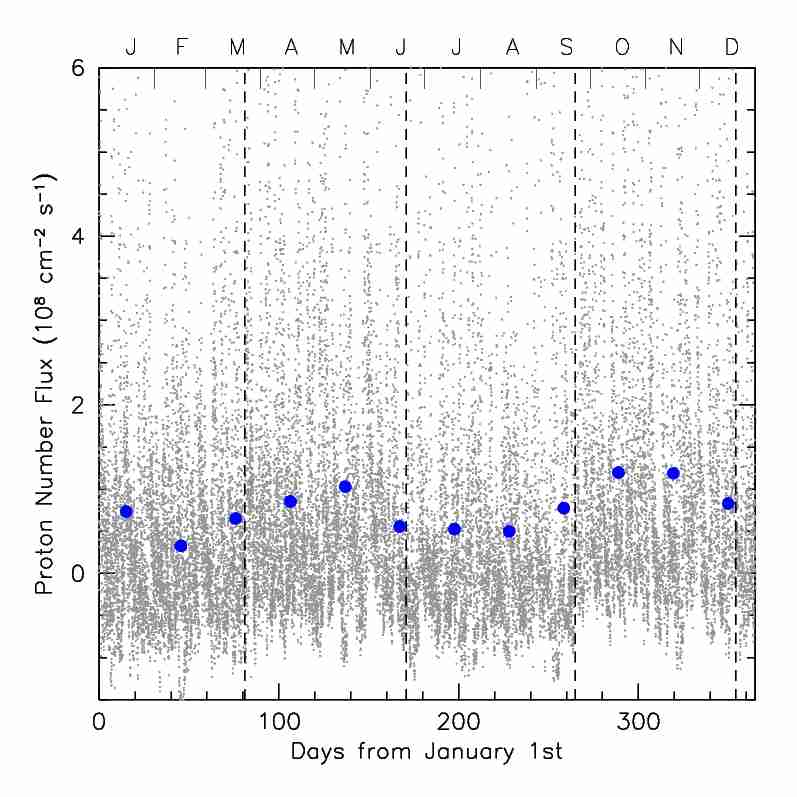}
\caption{\label{fig:seasonprot}Seasonal variation of the MTOF/CELIAS proton 
number flux. Only data from 2004-01-01 to 2007-01-01 are plotted. The
filled circles trace the monthly averages. No correction for the
11-years solar cycle has been applied.}
\end{figure}

In principle, since the solar rotation axis is inclined by about 7.1
degrees with respect to the ecliptic (Balthasar, Stark \& W\"ohl
\cite{balthasar}), and because the solar wind flux is known to be
higher from the poles than the equator, a modulation of the solar wind
as seen from Earth is indeed expected. More precisely, the proton flux
should be maximum around March 5 and September 5 when Earth lies at
its highest/lowest heliographic latitude, respectively. For the same
reason, the proton flux should be minimum around January 5 and June 5.
This prediction can be easily compared to the real data using the
CELIAS/MTOF measurements and looking at their behavior as a function
of time elapsed since the beginning of the year. The result is shown
in Fig.~\ref{fig:seasonprot}, which was produced using data obtained
towards solar minimum (January 2004 to January 2007), for a total of
about 28,000 data points. Indeed the PNF shows a SAO, with maxima in
April and November, i.e. significantly shifted in time with respect to
the epochs of maximum/minimum heliographic Earth's latitude.
Remarkably, the SAO observed in the PNF appears to be in phase with
the SAO detected both in the broad band data (Sec.~\ref{sec:season})
and in the emission features (Sec.~\ref{sec:spectra}).  Interestingly,
a similar plot for the 10.2 cm radio flux does not show any clear
trace of a SAO. Also, the evidence of a SAO in the PNF for the years
around the maximum of solar cycle n.~23 becomes weaker. A possible
explanation is that the more frequent and probably more stochastic
PEs tend to dominate over the smoother SAO when the sun is
more active.

\begin{figure}
\centering
\includegraphics[width=9cm]{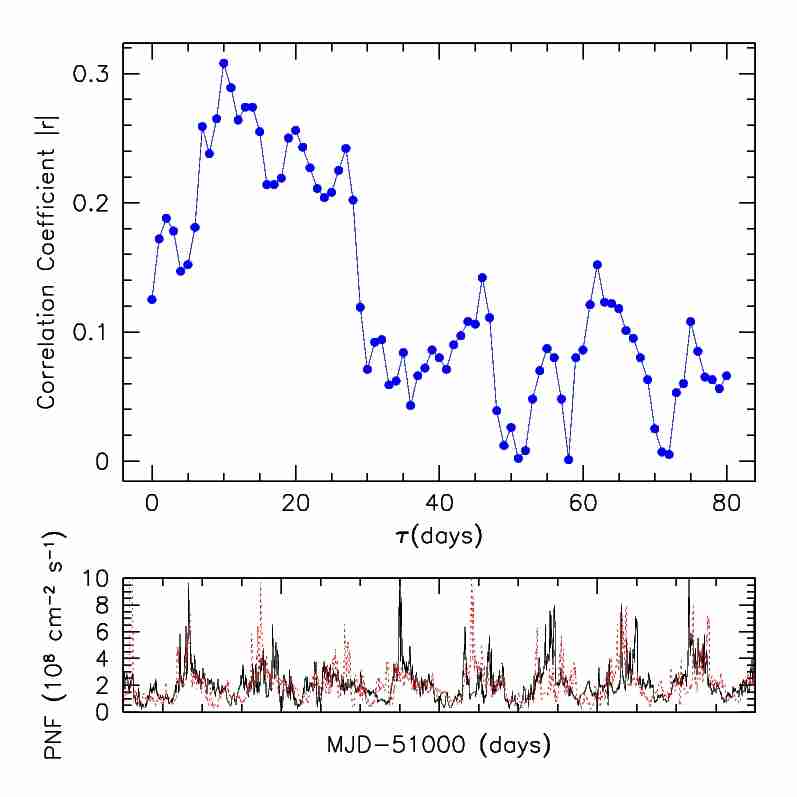}
\caption{\label{fig:sohoanal}Upper panel: Linear correlation coefficient 
between dark time $V$ sky brightness and PNF as a function of time
delay $\tau$, computed for $\Delta \tau$=10 days. Lower panel: example
of PNF periodicity. The dotted curve is a replica of the original data
shifted by 9.1 days.}
\end{figure}

Besides the smooth, long-term variation possibly produced by the
modulation of the average proton fluence, it is reasonable to think
that isolated PEs might be the cause of sporadic nightglow
enhancements. To explore this possibility, I have run a correlation
analysis similar to the one described in Sec.~\ref{sec:timescale} for
the solar radio flux. The results are in general rather noisy and show
that the correlation with PNF is always low ($|r|\leq$0.3), both for
short ($\Delta t$=0.5 days) and long ($\Delta t$=30 days) time
windows. An example for the $V$ passband is presented in
Fig.~\ref{fig:sohoanal} (upper panel) for an averaging window $\Delta
t$=10 days. The correlation coefficient reaches a peak $\sim$0.3 for
$\tau\simeq$10 days, it decreases and it suddenly drops for $\tau>$27
days. Even though the result is not really convincing, it might
indicate that the response of the night sky brightness to PEs takes
place with some time delay and that what matters is the proton flux
behavior during the last solar rotation. It must be noticed that the
PNF shows a rather marked recurrence, with a period of 9.1 days, that
coincides with one third of the solar rotation period (see
Fig.~\ref{fig:sohoanal}, lower panel). Similarly to what happens for
the radio data (see Sec.~\ref{sec:timescale}), this probably creates
spurious correlation peaks. Interestingly, shorter-timescale periodic
variations in the geomagnetic activity have been detected for 13.3,
9.1 and 6.9 days (see for instance Hauska, Abdel-Wahab \& Dyring
\cite{hauska}).

Since the typical speed of solar wind is $\sim$450 km s$^{-1}$, the
swarms of particles released during the PEs reach the Earth about one
hour after being detected by SOHO. Therefore, the delay in the
reaction is completely due to processes taking place within the
Earth's atmosphere.

As a last check for intermediate timescale periodic variations, I have
investigated the correlation with the Moon motion. In fact, it has
been suggested that atmospheric tides might induce recurrent
oscillations in the night sky brightness (see Chamberlain
\cite{chamberlain} and references therein). For this purpose, I have
investigated possible links between the broad band, dark time
measurements and lunar phase or lunar hour angle. No significant
correlation has been found, in agreement with a similar analysis run
by Mattila et al. (\cite{attila}).

\section{\label{sec:concl}Conclusions}

In this paper I have presented a photometric and spectroscopic
analysis of the optical night sky emission at Cerro Paranal in the
time interval April 2001 - January 2007. The main conclusions of this
work can be summarized as follows;

\begin{itemize}

\item The UBVRI night sky brightness is well correlated with solar activity.
The correlation is maximum in $U$ and minimum in $I$.

\item The excursion between sunspot maximum and minimum of solar cycle n.~23   
is 0.6, 0.3, 0.3, 0.4 and 0.2 mag arcsec$^{-2}$ in $U, B, V, R$ and
$I$ passbands, respectively.

\item There are indications that the effects of solar activity on nightglow
are not identical across different solar cycles.

\item The reaction of the nightglow to the solar variations seems to take
place with a time delay of the order of 2-3 weeks.

\item The $R$ night sky brightness seems to react with a much shorter delay,
of the order of a few days.

\item $V, R$ and $I$ measurements show a clear semi-annual oscillation (SAO),
with a typical peak-to-peak variation of $\sim$0.5 mag arcsec$^{-2}$.
For the $B$ passband this oscillation is, if present, much smaller.

\item Maxima and minima of the SAO are out of phase with respect to the
Equinoxes and Solstices.

\item All main emission features show a SAO, very similar to the
well known seasonal oscillation of the Na~I D doublet.

\item $[OI]$ 5577 and NI~5200\AA\/ show the strongest correlation with 
solar activity. For [OI]~5577, the maximum correlation is found for a
time delay of 15 days.

\item $[OI]$ 6300\AA and NI~5200\AA\/ show a very tight mutual correlation.
Nevertheless, [OI]6300 displays a weaker correlation with solar
activity.

\item Flux variations in the OH bands are very strongly correlated with
each other and do not show any correlation with solar activity.

\item The main emission features, both atomic and molecular, show smooth
flux variations on time scales of hours.

\item A preliminary and exploratory analysis of the possible connection
between nightglow and flux of charged particles from the Sun has shown
that there is a weak correlation. 

\item The night sky emission seems to react, with a delay of about 10 days,
to variations in the proton number flux.

\item No correlation is found between the dark time, broad band night sky 
brightness and moon phase or moon hour angle.

\end{itemize}

Future investigations, with even larger databases, will have to
address the possible relations with other solar phenomena, like
Coronal Mass Ejections, Flares and Solar Proton Events, in an attempt
to connect the observed short timescale variations of the nightglow
with space weather.

\begin{acknowledgements}

I am grateful to K. Krisciunas, for suggesting me to investigate the
time scales of the night sky brightness dependency on the solar
activity.  I also wish to thank R. Mignani, S. M\"ohler and the ESO
Quality Control Group for the support received during this
work. Finally, I am grateful to N. Castro and M. Garcia for reporting
the weird case of November 8, 2004. This paper is based on archival
data obtained with ESO Telescopes at Paranal Observatory.

\end{acknowledgements}

\end{document}